\documentclass[11pt,a4paper]{article}

\usepackage{jheppub}

\usepackage{amssymb,amsmath}
\usepackage{multicol,bbm}
\usepackage{bm}
\usepackage{multirow}
\usepackage{xcolor}
\usepackage{mathrsfs}
\usepackage{graphicx}
\usepackage{slashed}
\usepackage{epstopdf}
\usepackage[utf8]{inputenc}

\def\beq{\begin{equation}}
\def\eeq#1{\label{#1}\end{equation}}
\def\eeqn{\end{equation}}
\def\beqa{\begin{eqnarray}}
\def\eeqa#1{\label{#1}\end{eqnarray}}
\def\eeqan{\end{eqnarray}}

\def\leqn#1{(\ref{#1})}

\def\tp{t^\prime}
\def\mtp{m_{t^\prime}}
\def\Ytp{Y_{\tp\tp h}}
\def\F{{\cal F}}
\def\mT{m_T}
\def\YT{Y_{TT h}}
\def\T{T}

\def\stacksymbols #1#2#3#4{\def\theguybelow{#2}
	\def\vp{\lower#3pt}
	\def\sp{\baselineskip0pt\lineskip#4pt}
	\mathrel{\mathpalette\intermediary#1}}

\def\intermediary#1#2{\vp\vbox{\sp
		\everycr={}\tabskip0pt
		\halign{$\mathsurround0pt#1\hfil##\hfil$\crcr#2\crcr
			\theguybelow\crcr}}}

\def\gsim{\stacksymbols{>}{\sim}{2.5}{.2}}

\newcommand{\bspace}{\!\!\!\!}
\def\met{\mbox{$E{\bspace}/_{T}$}}

\newcommand{\comment}[1]{}

\newcommand{\TeV}{~\textrm{TeV}}
\newcommand{\GeV}{~\textrm{GeV}}

\newcommand{\fig}[1]{Fig.~\ref{fig:#1}}

\newcommand{\tab}[1]{Table~\ref{tab:#1}}

\newcommand{\bea}{\begin{eqnarray}}
\newcommand{\eea}{\end{eqnarray}}

\title{Naturalness Sum Rules and Their Collider Tests}

\author[a]{Csaba Cs\'aki,}
\author[b]{Felipe Ferreira De Freitas,}
\author[a,b,c]{Li Huang,}
\author[b,c]{Teng Ma,}
\author[a]{Maxim Perelstein,}
\author[b,c,d,e]{Jing Shu}

\emailAdd{csaki@cornell.edu, huangli@itp.ac.cn, mat@itp.ac.cn,mp325@cornell.edu,jshu@itp.ac.cn}

\affiliation[a]{Laboratory for Elementary Particle Physics, Cornell University, Ithaca, NY 14853, USA}
\affiliation[b]{CAS Key Laboratory of Theoretical Physics, Institute of Theoretical Physics,
Chinese Academy of Sciences, Beijing 100190, China}
\affiliation[c]{School of Physical Sciences, University of Chinese Academy of Sciences, Beijing 100049, P. R. China}
\affiliation[d]{CAS Center for Excellence in Particle Physics, Beijing 100049, China}
\affiliation[e]{Center for High Energy Physics, Peking University, Beijing 100871, China}

\abstract{We present the most general sum rules reflecting the cancellation of ultraviolet divergences in the Higgs potential in weakly-coupled, natural extensions of the Standard Model. There is a separate sum rule for the cancellation of the quadratic and logarithmic divergences, and their forms depend on whether the divergences are canceled by same-spin or opposite-spin partners. These sum rules can be applied to mass eigenstates and conveniently used for direct collider tests of naturalness. We study in detail the feasibility of testing these sum rules in the top sector at a future $100\TeV$ proton collider within two benchmark models, the Little Higgs (LH) and the Maximally Symmetric Composite Higgs (MSCH). We show how the two ingredients of the sum rules, the top partner masses and their Yukawa couplings to the Higgs, can be measured with sufficient accuracy to provide a highly non-trivial quantitative test of the sum rules. In particular, we study observables sensitive to the sign of the top partner Yukawa, which is crucial for verifying the sum rules but is notoriously difficult to measure. We demonstrate that in the benchmark models under study, a statistically significant discrimination between the two possible signs of each Yukawa will be feasible with a 30 ab$^{-1}$ data set at $100\TeV$.  
}

\begin{document}

\maketitle	

\section{Introduction} 
\label{sec:introduction}

Naturalness of electroweak symmetry breaking has long been recognized as the key issue that motivates extending the Standard Model (SM), with additional particles and interactions required to appear around the TeV scale in a natural theory. Experimental data, starting with precision electroweak tests in the 1990's and culminating with the discovery of the Higgs boson in 2012, confirmed the SM picture of electroweak symmetry breaking, and provided strong indication that theory must remain weakly coupled at the TeV scale. To be natural, weakly coupled theories require the presence of one or more new particles at the TeV scale, whose quantum-mechanical contributions to the Higgs potential must cancel the dominant contributions from the SM. At the very least, a natural, weakly-coupled theory with a cutoff of at least around 10 TeV requires that one-loop quadratically divergent contributions to the Higgs mass parameter from the top and electroweak gauge sectors be cancelled. In theoretically motivated constructions, the cancellation is enforced by a symmetry structure of the theory. Particles responsible for canceling the top and gauge contributions to the Higgs mass are called the ``top partners" and ``gauge partners", and their properties are related by symmetry to those of their SM counterparts, the top quark and the elecvtroweak gauge bosons. In particular, the couplings of the top/gauge partners to the Higgs boson are tightly constrained by the requirement of the divergence cancellation. The first goal of this paper is to provide a simple and general form of these constraints, which we call {\it naturalness sum rules}. Our derivation of the sum rules is very general and does not assume any specific form of physics Beyond the Standard Model (BSM). In addition to the ``leading" sum rule reflecting the cancellation of quadratic divergences, we also provide a second, ``sub-leading" sum rule which holds in theories where logarithmic divergences are also canceled. Such theories, exemplified by the recently proposed Maximally Symmetric Composite Higgs (MSCH) model ~\cite{Csaki:2017cep,Csaki:2018zzf}, are theoretically attractive, since the amount of fine-tuning in electroweak symmetry breaking is reduced compared to more traditional composite Higgs models (CHMs). \footnote{For reviews of CHM's see~\cite{Contino:2010rs,Bellazzini:2014yua,Panico:2015jxa,Csaki:2015hcd,Csaki:2016kln}.}

The naturalness sum rules in the top sector are expressed in terms of the masses of the top quark and its partner(s), and their Yukawa couplings to the Higgs boson.  If top partners are discovered, measuring their masses and Yukawa couplings can provide a direct experimental test of the sum rules. Such a test would establish whether the observed particles indeed restore naturalness, and whether the divergence cancellation occurs only at the quadratic level or at the level of log divergences as well. LHC searches for top partners have placed strong limits on these particles' masses, in all but rather exotic decay scenarios.\footnote{A well-known exception occurs in models of neutral naturalness~\cite{Chacko:2005pe,Csaki:2017jby}, where the top partners do not carry SM color. While our sum rules do apply in those theories, we are not aware of any possibility to test them experimentally in that case, and we will not discuss this case further.} While future LHC runs may well discover top partners, it is highly unlikely that they will collect samples of top partners large enough for a statistically significant test of the sum rules, given the current constraints. On the other hand, a 100 TeV proton-proton collider, such as the SppC~\cite{CEPC-SPPCStudyGroup:2015csa} proposed in China, or CERN's FCC-hh~\cite{Mangano:2017tke}, may provide an opportunity to explore the top partners in detail. In this paper, we demonstrate with a detailed Monte Carlo study that the naturalness sum rules can be tested at a 100 TeV collider, at a few-\% level, for two representative benchmark models of new physics: a Little Higgs (LH) model~\cite{ArkaniHamed:2002qy,Schmaltz:2005ky,Perelstein:2005ka}, and the MSCH model~\cite{Csaki:2017cep}. Crucially, we show that the {\it signs} of the top partner Yukawa couplings, and not just their magnitude, can be determined experimentally. This is especially important because of the role played by these signs in divergence cancellation.

Before proceeding, let us briefly comment on a few related earlier studies. The idea of sum rules that follow directly from divergence cancellation in natural theories, and relate experimentally observable quantities, has been explored before in a more model-specific context. In the run-up to the LHC, sum rules in the Little Higgs~\cite{Perelstein:2003wd} and supersymmetric (SUSY) models~\cite{Blanke:2010cm,Perelstein:2012fc} have been proposed. Since direct measurement of the top partners' Yukawa couplings is difficult at the LHC, these sum rules involved other, more easily measurable quantities, but as a result, they were model-dependent. Prospects of a direct measurement of the top partner ``Yukawa" in SUSY were studied in~\cite{Chang:2011jk}. More recently, Ref.~\cite{Chen:2017dwb} derived a sum rule involving the top partner Yukawas. This derivation was more general than in previous works, applying in a broad class of Little Higgs-type models with quadratic divergence cancellation. Within this class of models, the first of our sum rules is identical to that of~\cite{Chen:2017dwb}, but our derivation of it is both simpler and even more broadly applicable. Moreover, Ref.~\cite{Chen:2017dwb} performed a phenomenological study to demonstrate that the top partner Yukawa coupling can be measured at the 100 TeV collider, proposing to infer it from the rate of the associated Higgs-top partner process $pp\to \tp\tp h$. The collider study in our paper is complementary, since we focus on somewhat rarer processes, such as the Higgs-{\it single} top  partner production, $pp\to q\tp h$, which give access to the sign of the top partner Yukawa as well as its magnitude.           

The rest of the paper is organized as follows. In section~\ref{sec:setup}, we derive the general naturalness sum rules of the fermion sector, including the 
model-independent proof, and present generalizations to the gauge sector and for oppoiste spin partners. We also discuss two benchmark model examples, LH and MSCH models. In section~\ref{sec:collideroutline}, we outline the strategy to test the sum rules at a hadron collider. In section~\ref{sec:colliderstudy}, we present a detailed Monte Carlo study which demonstrates that the proposed sum rules can be tested at the 100 TeV collider. 
We conclude with a brief discussion of our findings in section~\ref{sec:discussion}.  
The Appendix contains further details on the derivation of sum rules, jet tagging, and the Random Forest algorithm used in our collider study.

\section{Naturalness Sum Rules}
\label{sec:setup}

In this section we establish very general sum rules that any theory without quadratic (or log) divergences have to satisfy. These sum rules provide the essential underlying mechanism for the solution of the hierarchy problem: testing them in high energy collider experiments should be of highest priority. Our considerations in this section will be very general and encompass all models regularly considered where top partners provide the cancellation of the quadratic divergences. 
We first focus on the case where the Higgs is a pseudo-Nambu-Goldstone boson (pNGB) of a global symmetry $G$ spontaneously broken to its subgroup $H$, as it happens in all  CHMs and LH. In these kinds of models, the quadratic UV sensitivity of the Higgs potential due to top or electroweak gauge bosons loops are cancelled by their corresponding  partners with same spin. We show that if the Higgs potential from the gauge and top sectors is natural, the couplings of the standard model fields and their partners to the physical Higgs boson should satisfy some sum rules. We first derive the most general sum rules for the Yukawa couplings of the top sector (including the top partners) and then briefly show the corresponding sum rules in the gauge sector. The detailed derivations for the gauge sum rules are presented in App.~\ref{App:gaugesumrules}. Then we present the general sum rules for the SUSY like models in which the Higgs is elementary and its divergences from top and gauge bosons loops are cancelled by  their partners with opposite spin while more details can be found  in App.~\ref{App:supersumrules}.

\subsection{Sum rules for the top sector: cancellation of quadratic divergences}

In unitary gauge, the mass matrix of the fermions in top sector in the background of Higgs $H$ can be parametrized as  
\beq
M_F=M_{F0} + f\, M(\frac{H}{f}),
\eeq
where $f$ is the global symmetry breaking scale, $M_{F0}$ and $M(H/f)$ are respectively the mass matrix independent and dependent on the Higgs field (hence satisfying the condition $M(0) =0$). The quadratically divergent contribution to the one loop Coleman-Weinberg potential from the top sector is proportional to 
\bea
V(H) \sim \mbox{Tr}[M_F.M_F^\dagger] \Lambda^2, 
\label{eq:quaddiv}
\eea  
where $\Lambda$ is the cutoff scale.
If there are no quadratically divergent contributions to the Higgs mass then $V(H)$ in (\ref{eq:quaddiv}) should not contain a quadratic term in $H$.
Thus we get the main condition for the cancelation of the quadratic divergences    
\bea \label{eq:naturalness1}
\frac{\partial^2   }{\partial H^2} \mbox{Tr}[M_F.M_F^\dagger]\vert_{H=0} =0
\eea
We can now turn this main condition into a sum rule involving the masses and the Yukawa couplings of the fermions. We will show that the cancelation of the quadratic divergences implies the vanishing of the tadpoles of the physical Higgs in the quadratically divergent part of the potential after electroweak symmetry breaking (EWSB). Assume the pNGB Higgs gets a VEV $\langle H \rangle =\tilde{v}$, and denote the fluctuating field above this VEV as $h$,
 \bea
 H = \tilde{v} + h.
 \eea  
Notice that the VEV of pNGB Higgs is not exactly equal to the EWSB scale $v=246$ GeV. To avoid confusion, in this paper we always follow the convention that the nonlinear sigma field describing the pNGBs $\pi^{\hat{a}}$ in coset space $G/H$  is parametrized as 
\bea \label{eq:sigma}
U=\mbox{Exp}\Big(\frac{i \sqrt{2} \pi^{\hat{a}} T^{\hat{a}} }{f} \Big),        
\eea
where $T^{\hat{a}}$ are the broken generators with normalization $\mbox{Tr}[T^{\hat{a}} T^{\hat{b}}] = \delta^{\hat{a} \hat{b}}$. Using these conventions, the full relation between the pNGB Higgs VEV and physical EWSB scale is 
\bea
v=f \sin \frac{\tilde{v}}{f}. 
\eea 
For the successful EWSB region $\frac{\tilde{v}}{f} \ll 1$, the EWSB scale is approximately equal to pNGB Higgs VEV, $v \approx \tilde{v} + \mathcal{O}(\tilde{v}^3/f^2)$. We will be neglecting other corrections of this order and hence we will not be making a distinction between $v$ and $\tilde{v}$.    
     
We find that Higgs boson $h$ tadpole in $V(H)$ after EWSB is proportional to the same mass terms that control the quadratic divergences before EWSB,     
\bea \label{eq:natural_equation}
&&\frac{\partial \mbox{Tr}[M_F. M_F^\dagger]}{\partial h}  \vert_{h=0}  \nonumber \\
&=& \big( \frac{\partial (H/f)}{\partial h}|_{h=0} \frac{\partial \mbox{Tr}[M_F. M_F^\dagger]}{\partial(H/f)}  \vert_{H=0}  + \frac{\partial (H/f)^2}{\partial h}|_{h=0} \frac{\partial^2 \mbox{Tr}[M_F. M_F^\dagger]}{2\partial^2(H/f)}  \vert_{H=0}  +\mathcal{O}\left(\tilde{v}^2/f^2 \right) \big) \nonumber \\
&=& \frac{\partial (H/f)^2}{\partial h}|_{h=0} \frac{\partial^2 \mbox{Tr}[M_F. M_F^\dagger]}{2\partial^2(H/f)}  \vert_{H=0}  +\mathcal{O}\left(\tilde{v}^2/f^2 \right) 
\eea
In this derivation we are first Taylor expanding $M_F M_F^\dagger = M_F M_F^\dagger|_{H=0} + H \frac{\partial}{\partial H} M_F M_F^\dagger|_{H=0} +\frac{1}{2} H^2 \frac{\partial^2}{\partial^2 H} M_F M_F^\dagger|_{H=0} +\ldots $ where each of the Taylor series coefficients are constants (independent of $H$ or $h$). The first term in the second line is $H$ tadpole in $V(H)$  and hence must be vanishing by SU(2) gauge invariance:
\bea
\frac{\partial \mbox{Tr}[M_F. M_F^\dagger]}{\partial(H/f)}\vert_{H=0} =0.
\eea  
while the second term is directly the coefficient of the quadratically divergent Higgs mass term.
Thus we find that if the Higgs mass term is not quadratically divergent, a sum rule for the terms linear in $h$ in the Higgs potential must be satisfied up to $\mathcal{O}(\tilde{v}^2/f^2)$
\bea \label{eq:natural_condition}
\frac{\partial \mbox{Tr}[M_F. M_F^\dagger]}{\partial h}  \vert_{h=0}  =0+\mathcal{O}(\tilde{v}^2/f^2).
\eea  
This sum rule can now be easily converted into one that contains the physical masses and Yukawa couplings. 
 We know that $M_{F}\vert_{h=0}$ is the fermion mass matrix  (after EWSB) and $(\frac{\partial}{\partial h} M_F)\vert_{h=0} $ is the matrix of physical Yukawa couplings. We can diagonize the fermion mass matrix up to order $\mathcal{O}(\tilde{v}^2/f^2)$ with a bi-unitary transformation:
\bea
L^\dagger M_F\vert_{h=0} R =M_D \quad L^\dagger(\frac{\partial}{\partial h} M_F)\vert_{h=0} R  =Y_M,
\eea
where $L$ and $R$ are the rotations of the left- and right-handed fermions, $M_D$ is the diagonal mass matrix and $Y_M$ is the Yukawa matrix for the mass eigenstates. We can now apply the same rotations to our naturalness condition (\ref{eq:natural_condition})  
\bea \label{eq:natural_eq2}
&&\frac{\partial \mbox{Tr}[M_F. M_F^\dagger]}{\partial h}\vert_{h=0}   = \mbox{Tr}[Y_M.M_D ^\dagger + M_D.Y_M^\dagger ] =0 +\mathcal{O}(\tilde{v}^2/f^2).  \nonumber \\   
\eea  
If we assume that CP is conserved then the vanishing of the quadratically divergent corrections to the Higgs mass terms yield a simple sum rule for the physical Yukawa couplings and fermions masses   
\bea \label{eq:natural_equation1}
&&\mbox{Tr}[Y_M.M_D  ]  =0 +\mathcal{O}(\tilde{v}^2/f^2).  \nonumber \\
\eea

\subsection{Sum rules for the top sector: cancellation of the log divergences}

One can carry through the same analysis to obtain a sum rule resulting from the vanishing of the the logarithmically divergent contributions to the Higgs mass. The log divergence in the Coleman-Weinberg potential is proportional to $\mbox{Tr}\big[(M_F.M_F^\dagger)^2 \big]\mbox{log} \Lambda^2  $ so the condition for the absence of a logarithmic divergence in the Higgs mass is 
\bea \label{eq:naturalness}
\frac{\partial^2}{\partial H^2} \mbox{Tr}[(M_F.M_F^\dagger)^2]\vert_{H=0} =0
\eea    
Just as before, we can show that if the Higgs mass is not logarithmically  divergent and using that the tadpole of $H$ is always zero, the $h$ tadpole after EWSB must also be zero up to $\mathcal{O}(\tilde{v}^2/f^2)$
\bea \label{eq:natural_identity2}
&&\frac{\partial}{\partial h}\mbox{Tr}[(M_F.M^\dagger_F)^2]\vert_{h=0}  \nonumber \\
&=& \big( \frac{\partial (H/f)}{\partial h}|_{h=0} \frac{\partial \mbox{Tr}[(M_F. M_F^\dagger)^2]}{\partial(H/f)}  \vert_{H=0}  + \frac{\partial (H/f)^2}{\partial h}|_{h=0} \frac{\partial^2 \mbox{Tr}[(M_F. M_F^\dagger)^2]}{2\partial^2(H/f)}  \vert_{H=0}  +\mathcal{O}\left(\tilde{v}^2/f^2 \right) \big) \nonumber \\
&=& \frac{\partial (H/f)^2}{\partial h}|_{h=0} \frac{\partial^2 \mbox{Tr}[(M_F. M_F^\dagger)^2]}{2\partial^2(H/f)}  \vert_{H=0}  +\mathcal{O}\left(\tilde{v}^2/f^2 \right) \nonumber \\
&=&0+\mathcal{O}\left(\tilde{v}^2/f^2 \right) 
\eea

Diagonalizing the mass matrix  the identity in Eq.~(\ref{eq:natural_identity2}) becomes  
\bea
\frac{\partial}{\partial h}\mbox{Tr}[(M_F.M_F^\dagger)^2]\vert_{h=0}  
=2\mbox{Tr}[ Y_M M_D^\dagger M_D M_D ^\dagger  +Y_M^\dagger M_D.M_D^\dagger.M_D ] 
= 0 +\mathcal{O}(\tilde{v}^2/f^2)
\eea  
If CP is conserved, we obtain the sum rule the vanishing of the logarithmis divergence,  
\bea
\mbox{Tr}[ Y_M M_D^3]= 0 +\mathcal{O}(\tilde{v}^2/f^2).
\eea 

\subsection{Gauge sector sum rules}

Following the same logic, we can easily derive the naturalness sum rule in gauge sector. We obtain the following results (the details of the derivations are presented in App.~\ref{App:gaugesumrules}). The cancellation of the quadratic divergences implies the sum rule 
\bea
\mbox{Tr}[g_{VVh}] = 0 +\mathcal{O}(\tilde{v}^2/f^2). 
\eea    
where $g_{VVh}$ is the linear coupling of the physical Higgs boson to the physical (mass eigenstate) gauge bosons, which has mass dimension one. Similarly the cancellation of the log divergences will imply the sum rule 
\bea
\mbox{Tr}[g_{VVh}  M_V^2]=  0+ \mathcal{O}(\tilde{v}^2/f^2), 
\eea  
where $M_V$ is the diagonal mass matrix for the physical vector bosons. 

\subsection{Sum rules for opposite spin partners}

Our previous discussion of sum rules can be extended to the case where the divergences are cancelled by opposite spin partners as in supersymmetric (SUSY) theories. For completeness we briefly list these sum rules (while the more detailed derivation can be found in App.~(\ref{App:supersumrules})). The general sum rule for the cancellation of the quadratic divergences via opposite spin partners is
\bea \label{eq:SUSY_sumrule}
\text{Tr}[g_{SSh}] -2\text{Tr}[Y_M M_D^\dagger + M_D Y_M^\dagger ] +3  \text{Tr}[ g_{VVh} ]=0,  
\eea 
 where $g_{SSh} $ is the matrix of trilinear couplings of the SM Higgs boson to the physical (mass eigenstate) additional scalar particles. If CP is conserved, this general sum rule  will be simplified to
 \bea \label{eq:SUSY_sumrule_simple}
 \text{Tr}[g_{SSh}] -4\text{Tr}[Y_M M_D] +3  \text{Tr}[ g_{VVh} ]=0.
 \eea 
Note that in SUSY theories one generically has to deal with multiple Higgs doublets leading to several physical neutral Higgs bosons. Assuming that the quadratic divergences for all Higgs doublets are cancelled, the above sum rules should apply for every physical neutral Higgs boson separately. 

For example for the case of SUSY theories this sum rule can be directly translated to the separate sum rules for the top/stop and the gauge/gaugino/Higgs/ Higgsino sectors (assuming that the cancellation of the quadratic divergences in the top/stop sector are independent from the other sector):
 \begin{itemize}
\item Top sector/stop sector  
\end{itemize}
\bea \label{eq:SUSY_sumrule_top}
 \sum_i g_{\tilde{t}_i  \tilde{t}_i h}  -4y_{t} m_t =0,
 \eea  
 where the $g_{\tilde{t}_i  \tilde{t}_i h}$  are the trilinear physical Higgs-stop-stop  coupling to the  physical stops $\tilde{t}_i$  and $y_{t}$ is the top Yukawa coupling.
 
\begin{itemize}
 \item Gauge/gaugino/Higgs/Higgsino sector 
\end{itemize}
Generically the gauginos and Higgsinos will mix to form the neutralinos and charginos, and there will be a single sum rule for the entire sector: 
\bea
 && 4 \sum_{i} (y_{C_i^+C_i^- h } m_{C_i}+ y_{N_iN_i h } m_{N_i} ) -3 (g_{W^+W^- h} + g_{ZZ h})  
\nonumber \\ &&  -\sum_i ( g_{H^0_i H^0_i h} +g_{H^+_i H^-_i h} )- g_{hhh} =0
\eea   
where $ y_{C_i^+C_i^- h}$ and  $ y_{N_iN_i h}$  are the physical Yukawa coupling of  the charginos and neutralinos,    $m_{C_i}$ and  $m_{N_i}$ are the chargino and neutralino masses and   $g_{W^+W^- h}$ and  $g_{ZZ h}$  are Higgs trilinear couplings to the $W$ and $Z$ gauge bosons. In the second line $g_{H^0_i H^0_i h}$ and $g_{H^+_i H^-_i h}$ are the the trilinear couplings of the physical Higgs to new neutral and charged scalars and $g_{hhh}$ is the Higgs cubic self coupling.
 If there is a separate cancellation of the charged and the neutral sector there will be two separate sum rules involving only the terms corresponding to the charged/neutral states from the equation above. If there is no mixing between the Higgsinos and gauginos the sum rule may also split into separate sum rules for the gauge/gaugino and the Higgs/Higgsino sectors but generically there is just one overall sum rule that can be experimentally tested. 
 
It can be easily checked that all of these sum rules are satisfied within SUSY models.

\subsection{Explicit Examples examples for the sum rules and benchmark models}

Below we present our two benchmark models that we will be examining and show how the sume rules in the top sector are satisied here.

\subsubsection{Little Higgs}

Little Higgs (LH) models were the earliest successful pNGB Higgs models, which predict the presence of same spin top and gauge partners. In  LH models the quadratic divergences in Higgs potential are cancelled via the mechanism of collective symmetry breaking:  the Higgs shift symmetry is explicitly broken only when two separate terms in the Lagrangian are simultaneously turned on. If one of these terms is turned off  the Higgs shift symmetry will be restored  and no potential can be generated, hence the potential must be proportional to the product of these symmetry breaking couplings, softening the divergence and eliminating the quadratic divergence (while the potential is generically still log divergent). Since the Higgs quadratic divergences are cancelled, our first sum rule must be satisfied (while the second one for these models will not be satisfied). We will be chosing the top sector of the ``simplest Little Higgs model" (SLH) as one of our benchmark models. Here we first explain how the sum rule is analytically satisfied in this particular model while in the later part of the paper we show how this sum rule can be tested experimentally. 

The top sector of the SLH is quite general and representative of the top sector of many other LH models. In particular it is identical of the top sector of the ``littlest Higgs". The gauge group of the SLH model is SU(3), and there are two copies of the $SU(3)/SU(2)$ coset, where the gauged SU(3) is identified with the diagonal SU(3). The Higgs will be part of the non-linear field $U$ parametrized as in Eq.\ref{eq:sigma},
where the broken generators $T^{\hat{a}}$ are in direction $SU(3)/SU(2)$ and the Goldstone matrix $\pi^{\hat{a}} T^{\hat{a}}$ is given by (ignoring an additional singlet Goldstone that does not play a role in the cancellation of the quadratic diveregnces)
\bea
\pi^{\hat{a}} T^{\hat{a}}  =  \left( \begin{array}{ccc}
    0 & i \sigma_2 \tilde{H}^\ast \\ 
    -i \sigma_2 \tilde{H}^T & 0 \\   
  \end{array}  \right), 
\eea 
The linearly transforming Higgs triplet will then be given by 
\begin{equation}
{\cal H} = U \left( \begin{array}{c} 0 \\ 0 \\ f \end{array} \right)= \left( \begin{array}{c} 0 \\ i f \sin \frac{H}{f} \\ f \cos \frac{H}{f} \end{array} \right)\ . 
\end{equation}
where $H = \tilde{v} + h$.  
The top sector is obtained by extending the SM quark doublet to a full  SU(3) triplet by introducing an additional electroweak singlet Dirac fermion $T$. The left-handed component $T_L$ will be part of the SU(3) triplet 
\bea
\Psi_q = \left( \begin{array}{c}
    q_L \\ 
\tilde{T}_L 
  \end{array} \right)
\eea
while $T_R$ is an SU(3) singlet. The Yukawa couplings will then be written in the form 
\bea
\mathcal{L}_t =-\lambda_1 f \bar{\Psi}_q {\cal H} u_{3R} -\lambda_2 f \bar{\tilde{T}}_L \tilde{T}_R +h.c.,  
\eea   
where $u_{3R}$ is the usual RH top quark (which will mix with $T_R$ to form the actual SM $t_R$). 
It is easy to see that if we turn off one of Yukawa couplings $\lambda_{1,2}$, the NGBs can always be removed from the fermion sector, hence implementing collective symmetry breaking.  This also implies that the Higgs potential from the top sector should be proportional to $|\lambda_1\lambda_2|^2 f^4$ and only log divergent. The quadratic divergences from the usual top quark loop are cancelled by the loops involving the Dirac fermion $\tilde{T}$ which can be easily checked directly. The Fermion mass matrix in the electroweak (EW) basis $(t_{3L}, \tilde{T}_L)$ and $(u_{3R}, \tilde{T}_R)$ in the Higgs background is 
\bea
M_F =  -f \left(\begin{array}{cc}
    \lambda_1 \sin (\frac{H}{ f}) & 0 \\ 
    \lambda_1 \cos (\frac{H}{f})  &   \lambda_2 \\ 
  \end{array}\right).
\eea 
It is easy to see that the quadratically divergent term in the potential is independent of the  Higgs field 
\bea
 \mbox{Tr}[M_F.M_F^\dagger] \Lambda^2 =  f^2(\lambda_1^2 +\lambda_2^2)\Lambda^2 \ 
\eea
leading to the first sum rule. We can also explicitly check the validity of the first sum rule after EWSB.  After EWSB, the physical top and top partner masses are  
\bea
m_t =\frac{\lambda_1 \lambda_2 \tilde{v}}{\sqrt{\lambda_1^2 +\lambda_2^2}}+\mathcal{O}(\frac{\tilde{v}^2}{f^2}) \quad  m_T =- f \sqrt{\lambda_1^2 +\lambda_2^2} +\mathcal{O}(\frac{\tilde{v}^2}{f^2}).
\eea   
while their diagonal Yukawa couplings to the Higgs are  
\bea
y_{t} =-\frac{m_t}{\tilde{v}} +\mathcal{O}(\frac{\tilde{v}^2}{f^2}) \quad y_{T}= \frac{y_t^2 \tilde{v} }{m_T} +\mathcal{O}(\frac{\tilde{v}^2}{f^2}).
\eea
Thus we can see that in this model the first sum rule is indeed satisfied:
\beq
y_t m_t +  y_{T} m_T =0+\mathcal{O}(\frac{\tilde{v}^2}{f^2}). 
\eeq{LHsumrule}

\subsubsection{Maximally Symmetric Composite Higgs}

Here we briefly summarize the main general features of the MSCH model~\cite{Csaki:2017cep} and then discuss its top sector and in particular the corresponding sum rules in detail. 
The MSCH model is based on a symmetric coset space $G/H$ with an unusual unbroken global symmetry of the composite fermion sector called maximal symmetry. This unbroken flavor symmetry is the subgroup of the enhanced chiral global symmetry $G_L \times G_R$ of the composite fermions: the combinations of the generators of  $G_L \times G_R$ which leave the VEV $V$ that breaks $G$ to $H$ invariant. In the SM fermion effective Lagrangian (obtained after integrating out all the heavy composites), this symmetry will eliminate all the Higgs dependence in the effective kinetic terms and only allow the Higgs dependence only in the effective Yukawa couplings. As a consequence in models where the SM fermion masses are produced collectively via partial compositeness, the Higgs potential from effective Yukawa couplings will be automatically finite.  In this model, the Higgs potential is dominated by the top sector and has the minimal universal  tuning for successful EWSB. Since the Higgs potential from the top sector is finite,  we will have both sum rules satisfied, which we will explicitly display below. 

 We choose the minimal $SO(5)/SO(4)$ coset with custodial symmetry for the MSCH model. The EW gauge interaction is embedded in the $SO(4) \simeq SU(2)_L \times SU(2)_R$ subgroup. The NGBs from $SO(5)/SO(4)$ are also described by the nonlinear sigma field $U$ in Eq.\ref{eq:sigma}, where $T^{\hat{a}}$ is the broken generators in direction $SO(5)/SO(4)$ and $f$ is still the global symmetry breaking scale.  
A crucial element of the construction of the MSCH model is the presence of the Higgs-parity operator $V =\mbox{diag}(1,1,1,1,-1)$ which also can be identified with the Higgs VEV breaking $SO(5)\to SO(4)$. 

In the MSCH model it is assumed that the composite fermionic operator $\mathcal{O}$ mixing with the SM top is in the fundamental representation of $SO(5)$. Since the composite sector only has $SO(4)$ symmetry, this composite operator can be decomposed into a fourplet and a singlet top partner under this subgroup
\bea
 {\bf 5} ( \mathcal{O} )  \to {\bf 4}( \Psi_Q) + {\bf 1}( \Psi_S).
\eea
The explicit parametrization of these top partners is then:  
\bea
\Psi_Q = \frac{1}{\sqrt{2}}\left( \begin{array}{c}
i B -i X_{5/3} \\ 
B + X_{5/3} \\ 
i T + i X_{2/3} \\ 
-T +X_{2/3} \\
0
\end{array} \right) \quad \Psi_S =\left(  \begin{array}{c}
0 \\ 
0 \\ 
0 \\ 
0 \\ 
T_1
\end{array} \right).  
\eea

In order to mix with these composite fermions, the SM top doublet and singlet should also be embedded into the $\bf 5$ representation with the explicit form 
\bea
\Psi_{q_L} =\frac{1}{\sqrt{2}} \left( \begin{array}{c}
i b_L \\ 
b_L \\ 
i t_L \\ 
- t_L \\ 
0
\end{array} \right) \quad  \Psi_{t_R} = \left( \begin{array}{c}
0 \\ 
0 \\ 
0 \\ 
0\\ 
t_R
\end{array} \right). 
\eea
To realize maximal symmetry, the elementary-composite mixing terms should be fully $SO(5)$ invariant to ensure that these mixing terms are invariant under the fully enhanced chiral global symmetry $SO(5)_L \times SO(5)_R$.  The breaking of this symmetry appears only in the  composite top partner mass term which is twisted by $V$. This structure results in the chiral global symmetry broken to the maximal symmetry $SO(5)_{V^\prime}$:
\bea
g_L V g_R^\dagger =V\,, \quad  g_{L,R } \in SO(5)_{L,R}.
\eea       
As explained before this breaking pattern will ensure the finiteness of the Higgs potential. The explicit form of the Lagrangian of the top sector satisfying the requirements of maximal symmetry is   
\bea \label{eq:SO5mixing1} 
\mathcal{L}_{f} &=& \bar{q}_L i \slashed D q_L + \bar{t}_R i \slashed D t_R+ \bar{\Psi}_Q i \slashed \bigtriangledown  \Psi_Q + \bar{\Psi}_S i \slashed \bigtriangledown \Psi_S \nonumber \\ 
&-& \frac{1}{\sqrt{2}} \epsilon_{t}  f \bar{\Psi }_{t_R } U \Psi_{+L} - \epsilon_{q} f \bar{\Psi }_{q_L}  U \Psi_{+R} - M  \bar{\Psi}_{+L} V \Psi_{+R}   +h.c,            
\eea
where $\Psi_+ =\Psi_Q +\Psi_S$ is the full $5$ multiplet of $SO(5)$, $D_\mu$ is the covariant derivative for top and $\nabla_\mu =\partial_\mu -i E_\mu$, where $E_\mu$ is the $e$ symbol of CCWZ (but this is not important for the top sector contributions to the Higgs potential).  Maximal symmetry guarantees that the Higgs shift symmetry is collectively broken by the couplings $\epsilon_{t,q}$ and the mass $M$, hence the Higgs potential from the top sector loops must be proportional to $\sim (\epsilon_{t} \epsilon_{q})^2 f^4 M^2$. As expected  power counting shows that the degree of divergence of the Higgs potential  is negative so it is finite, free of quadratic and logarithm divergences. Now we can check our sum rules in this model explicitly. In unitary gauge, the mass matrix  for the top and the top partners in the background of the physical Higgs $H$ in the basis $(t, T, X_{2/3},T_1 )$ is 
\bea
\label{eq:mass}
&M_F = \left(
\begin{array}{cccc}
 0 & \epsilon_{q} f \cos^2 \left(\frac{{H}}{{2f }}\right)& \epsilon_{q}f \sin^2 \left(\frac{H}{{2f }}\right) & -\frac{\epsilon_{q} f \sin \left(\frac{H}{f} \right)}{\sqrt{2}} \\
 \frac{\epsilon_{t} f \sin \left(\frac{H}{ f}\right)}{2}  & M & 0 & 0 \\
 -\frac{ \epsilon_{t}f \sin \left(\frac{H}{f}\right)}{2} & 0 & M & 0 \\
 \frac{\epsilon_{t} f \cos \left(\frac{H}{f}\right)}{\sqrt{2}} & 0 & 0 & -M \\
\end{array}
\right).  
\eea
Due to the off-diagonal Yukawas, a further rotation is necessary to diagonalize the mass matrix after EWSB. We label the final (post-EWSB) mass eigenstates as $t_{1,2,3}^\prime$, with $m(\tp_1)\leq m(\tp_2)\leq m(\tp_3)$. 

It is easy to confirm that the Higgs quadratic and logarithmic  divergences vanish respectively since
\bea
\mbox{Tr}[M_F.M_F^\dagger] \Lambda^2 &=& ( 3 M^2 +\epsilon_{q}^2 +\frac{\epsilon_{t}^2}{2})\Lambda^2 \nonumber \\
\mbox{Tr}[(M_F.M_F^\dagger)^2] \mbox{log} \Lambda^2 &=& ( 3 M^4 + \epsilon_{q}^4 +\frac{\epsilon_{t}^4}{4} +(2\epsilon_{q}^2  +\epsilon_{t}^2   )M^2 )\mbox{log} \Lambda^2
\eea
are both independent of the Higgs background field $H$. 
After EWSB, the Yukawa coupling matrix can be extracted from $M_F$
\bea
Y =(\frac{\partial}{\partial H} M_F)\vert_{H=\tilde{v}}=\left(
\begin{array}{cccc}
 0 &- \frac{\epsilon_{q} \sin \left(\frac{{\tilde{v}}}{{f }}\right) }{2}& \frac{\epsilon_{q} \sin \left(\frac{{\tilde{v}}}{{f }}\right) }{2} & -\frac{\epsilon_{q} \cos \left(\frac{\tilde{v}}{f} \right)}{\sqrt{2}} \\
 \frac{\epsilon_{t} \cos \left(\frac{\tilde{v}}{ f}\right) }{2}   & 0 & 0 & 0 \\
 - \frac{\epsilon_{t} \cos \left(\frac{\tilde{v}}{ f}\right) }{2} & 0 & 0 & 0 \\
 -\frac{\epsilon_{t} \sin \left(\frac{\tilde{v}}{f}\right)}{\sqrt{2}} & 0 & 0 & 0 \\
\end{array}
\right).  
\eea                            
We can then explicitly check that  first and second sum rule are both exactly satisfied with these mass $M_F$ and Yukawa $Y$ matrices:
\beqa
\mbox{Tr}[Y_M.M_D^\dagger ] =\mbox{Tr}[Y.(M_F^\dagger) \vert_{H=\tilde{v}}] &=& 0 \,,\\  \mbox{Tr}[Y_M.M_D ^3 ] =\mbox{Tr}[Y.(M_F^\dagger. M_F.M_F^\dagger)\vert_{H=\tilde{v}}] &=& 0.  
\eeqa{MSCH_sumrules}

\section{Collider Tests of the Sum Rules: Outline}
\label{sec:collideroutline}
 
We have seen above that the cancellation of divergences via same-sign partners implies the presence of very simple and generic sum rules. The focus of the rest of the paper will be to establish how these sum rules for the top sector can be tested experimentally. Such a test requires measurements of the top partner mass $\mtp$, and its ``diagonal" coupling to the Higgs, $\Ytp$. (In this section, $\tp$ denotes a generic fermionic top partner. If multiple top partners contribute to divergence cancellation in a given model, the measurements outlined here must be performed separately for each one.) While the sum rules apply to any model with weakly-coupled top partners, details of the experimental strategies required to perform such measurements are model-dependent. To illustrate how such measurements can be performed in  concrete, well-motivated scenarios, we choose two models with fermionic top partners, namely the LH and MSCH models introduced above. 

Collider phenomenology of fermionic top partners has been studied by many authors~\cite{Li:2013xba}. The dominant production processes at a hadron collider are pair-production, $pp\to \tp\bar{t}^\prime$, and electroweak single production, $pp\to t^\prime q$. The produced top partners decay to SM final states\footnote{It is also possible for top partners to have non-SM decay channels, involving, for example, missing energy~\cite{Anandakrishnan:2015yfa}. We will not consider this possibiilty in this work.} $tZ$, $th$, and $bW$. Currently, searches for such final states at the LHC place bounds on the top partner mass of about $\mtp\gsim 1.3$~TeV~\cite{Aaboud:2017qpr,Aaboud:2017zfn,Sirunyan:2017ynj,Aaboud:2018xuw,Sirunyan:2018omb,Aaboud:2018pii}. For our study, we choose benchmark points in the parameter spaces of the LH and MSCH models that are consistent with these constraints, with top partner masses around 1.5~TeV. The benchmark points will be defined in Section~\ref{sec:BPs}. For these benchmark points, it is expected that future runs of the LHC will discover the top partners. However, we find that detailed measurements of the top partner properties needed for checking the sum rules require significantly larger data sets than what can be collected at the (HL-)LHC. Motivated by this, we study the potential of a future 100~TeV hadron collider, such as the SppC~\cite{CEPC-SPPCStudyGroup:2015csa} proposed in China, or CERN's FCC-hh~\cite{Mangano:2017tke}. 

The first ingredient of the sum rules that will be measured is the top partner mass $\mtp$. This mass can be reconstructed using fully visible decay channels, for example $tZ$ with a hadronic top (typically appearing as a top jet, given the high mass of the top partner) and leptonic $Z$. 
In the case of MSCH model, masses of all relevant top partners can be measured using this technique, with similar uncertainties. 

Measuring $\Ytp$ is more subtle. Ref.~\cite{Chen:2017dwb} proposed that this coupling can be determined from the rate of the process $pp\to \tp\bar{t}^\prime h$, which can be measurable at the 100 TeV collider. This method, however, has an important limitation: since the rate is proportional to $|\Ytp|^2$, it can only provide information about the {\it absolute value}, and not the {\it sign}, of the coupling. Relative signs between the terms in the sum rule are crucially important for cancellations to occur, and therefore it is necessary to test these signs experimentally. In this paper, we propose a novel observable that is sensitive to the sign of the top partner-Higgs coupling. The idea is simple. Consider a parton-level production process involving $\tp$, such that the leading matrix element receives two contributions: one with an odd number of $\Ytp$ vertices (for example, one), and the other one with an even number of $\Ytp$ vertices (for example, zero). The cross section of such a process includes the interference between the two contributions, which is sensitive to the sign of $\Ytp$. If the two contibutions are roughly similar in size, the interference term can constitute a sizeable fraction of the total cross section. Assuming all other relevant parameters are known, measuring this cross section provides information about both the magnitude and the sign of $\Ytp$. 

\begin{figure}[t]
	\begin{center}
		\includegraphics[width=0.32 \linewidth]{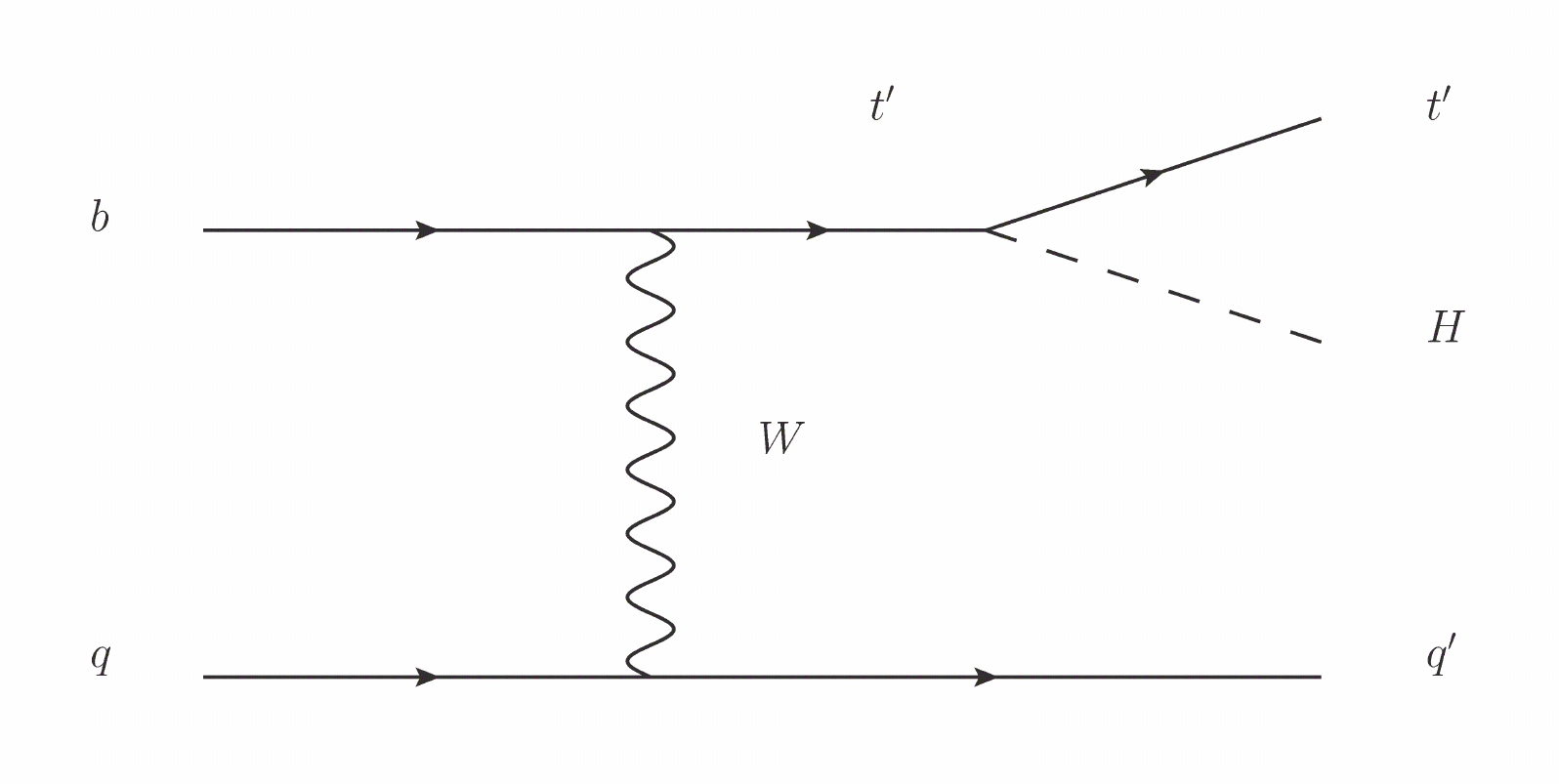}
		\includegraphics[width=0.32 \linewidth]{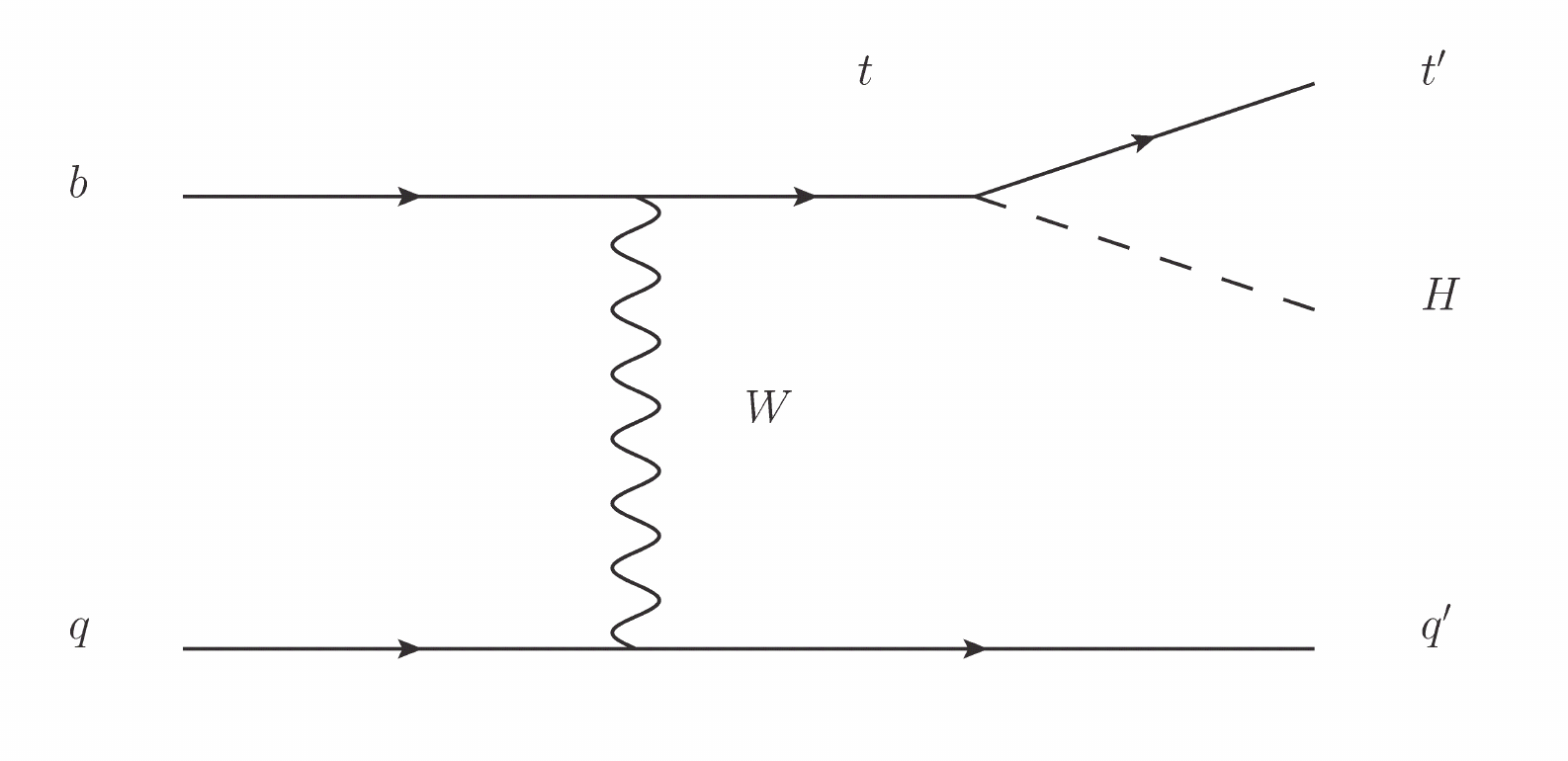}
		\includegraphics[width=0.32 \linewidth]{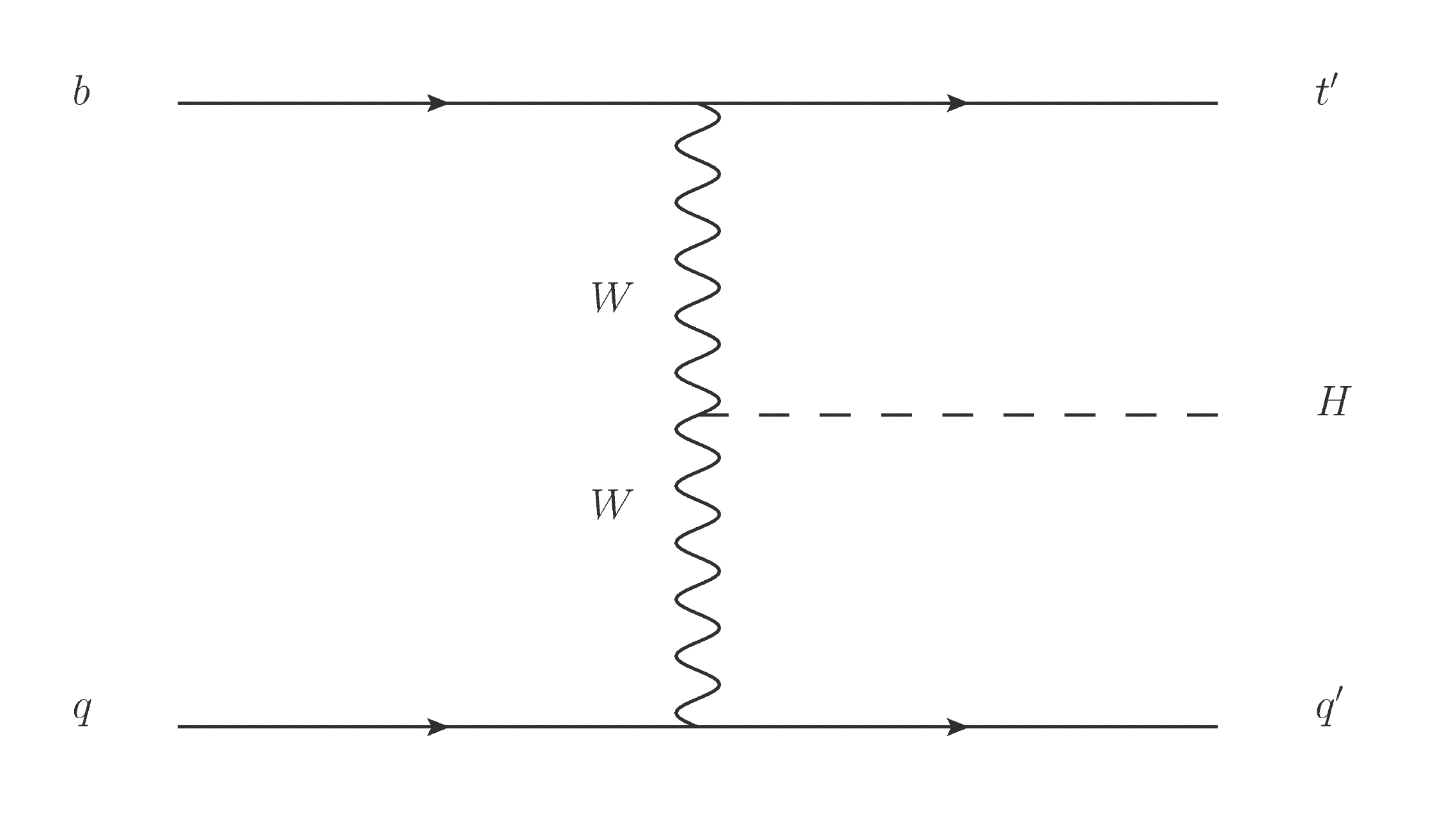}
	\end{center}
	\caption{Representative Feynman diagrams for single top partner production in association with the Higgs, $pp\rightarrow qht^{'}$. The diagram on the left changes sign when the sign of $\Ytp$ is flipped, while the other two diagrams do not.}
	\label{fig:FD_qhtp} 
\end{figure}

The simplest example of a process sensitive to the sign of $\Ytp$ is single top partner production in association with the Higgs, $pp\to q h \tp$. Representative Feynman diagrams for this process are shown in Fig.~\ref{fig:FD_qhtp}. The diagram on the left changes sign when the sign of $\Ytp$ is flipped, while the other two diagrams do not. If the off-diagonal $t\tp h$ coupling and the top-partner gauge couplings are measured independently, the total rate of this process can be used to infer $\Ytp$, incuding its sign. This approach is directly parallel to that proposed in Ref.~\cite{Farina:2012xp} to determine the sign of the SM top Yukawa coupling at the LHC using the process $pp\to qh t$. 

\begin{figure}[t!]
	\begin{center}
		\includegraphics[width=0.4 \linewidth]{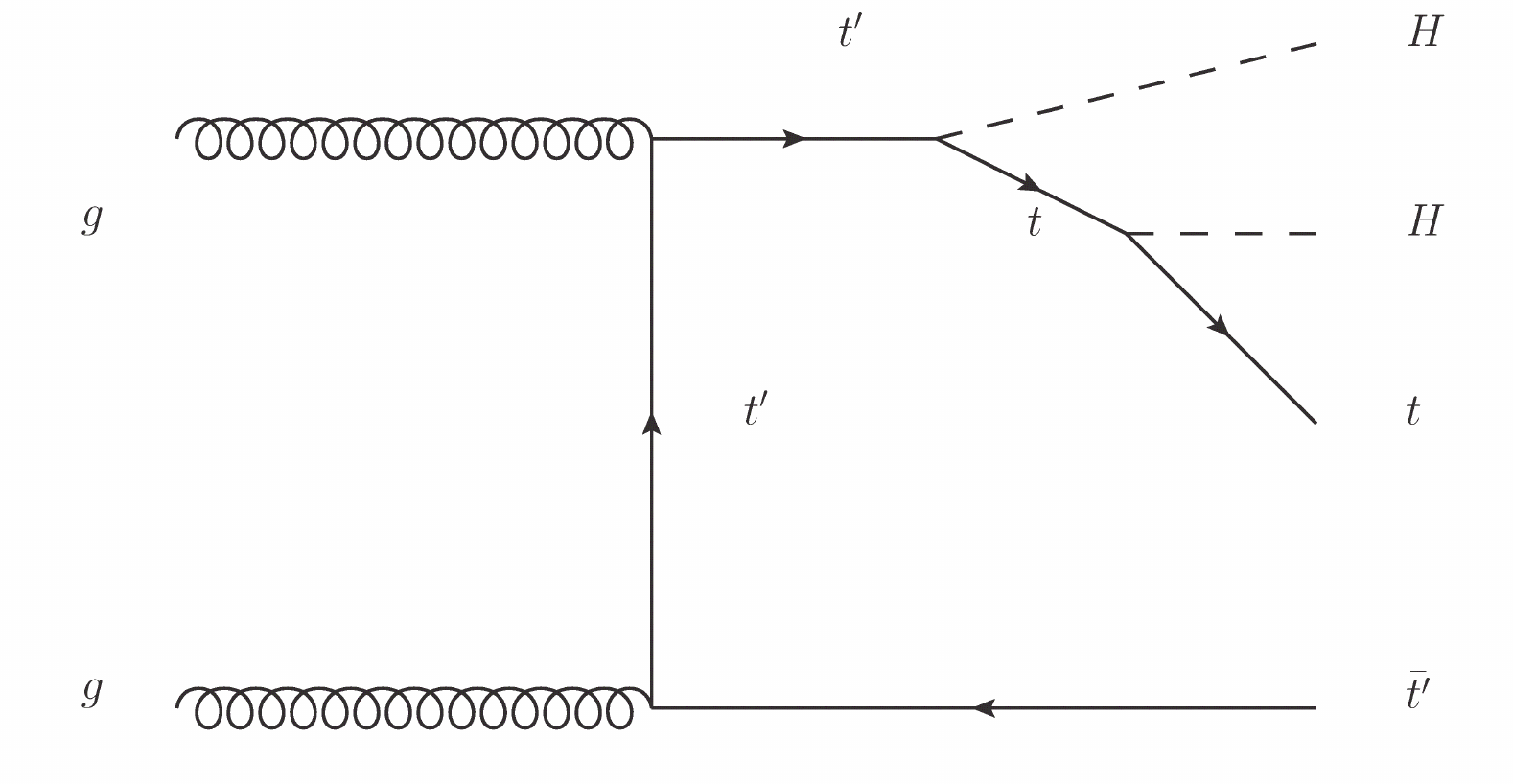}
		\includegraphics[width=0.4 \linewidth]{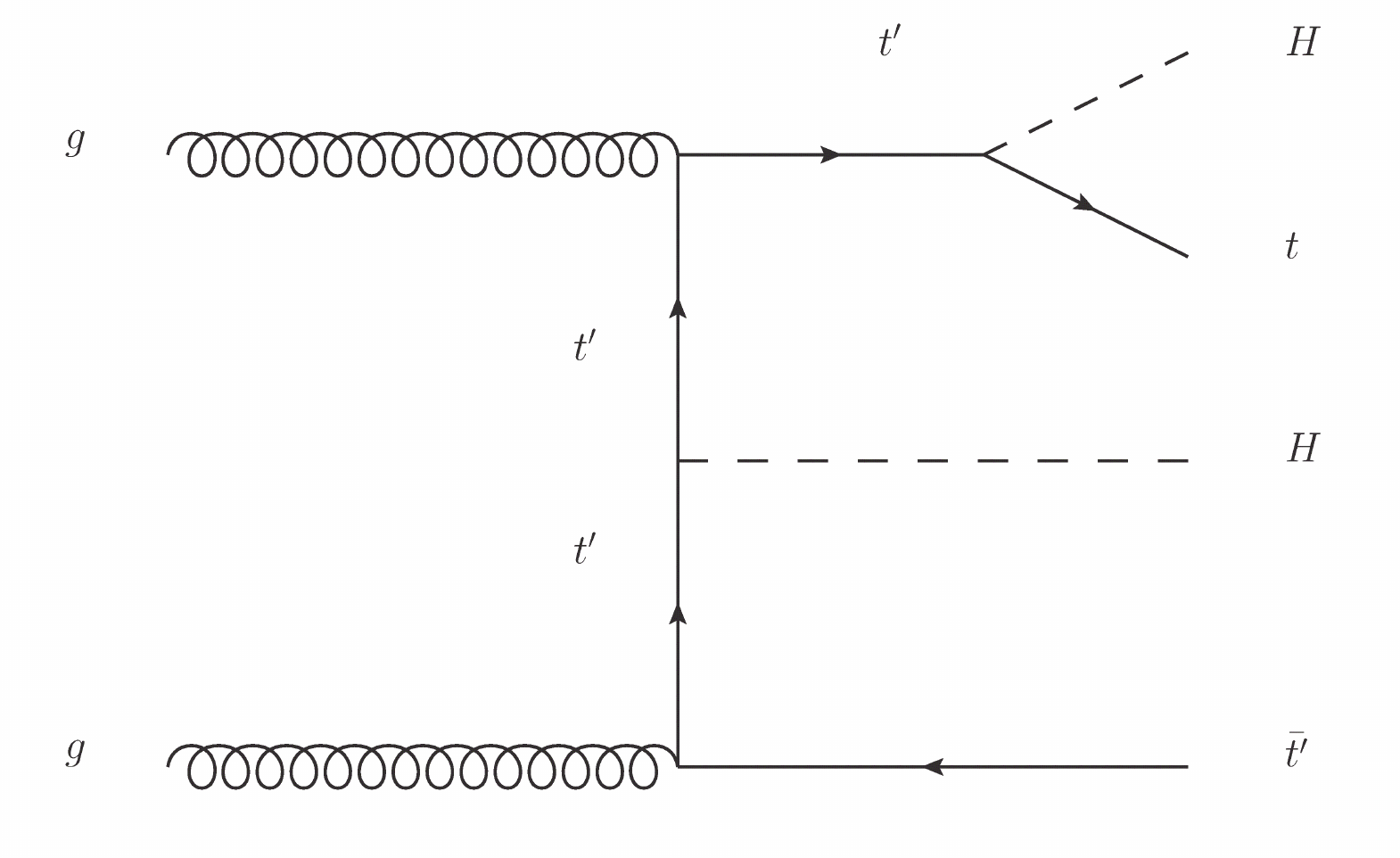}
	\end{center}
	\caption{Representative Feynman diagrams for the process $pp\rightarrow t^{'}thh$. The diagram on the right changes sign when the sign of $\Ytp$ is flipped, while the diagram on the left does not.}
	\label{fig:FD_tpthh} 
\end{figure}

Another interesting example is the process $pp\rightarrow t^{'}thh$. The dominant contributions to the matrix element of this process are illustrated in Fig.~\ref{fig:FD_tpthh}. The first contribution (left panel) is from top partner pair-production $pp\to\tp\tp$, followed by a three-body decay of one of the top partners, $\tp \to thh$, via a diagram with an off-shell top. The second contribution (right panel) is from top partner pair-production in association with the Higgs, $pp\to\tp\tp h$, followed by a two-body decay of one of the top partners, $\tp \to th$. (The second top partner decays in the same two-body channel in both cases, and this decay does not affect the discussion.) If the sign of $\Ytp$ is flipped, the first contribution is unaffected, while the second one changes sign. Since the two constributions are of the same order in perturbation theory, their interference makes up a significant part of the cross section, making this process promising for determining the sign of $\Ytp$ despite its relatively small rate.               

\section{Collider Tests of the Sum Rules: Feasibility Study}
\label{sec:colliderstudy}

To demonstrate that the proposed sum rules are useful in practice, we performed a detailed study of the relevant collider phenomenology. For this study, we chose two benchmark parameter points, one each in the LH and MSCH models. We then performed realistic Monte Carlo simulations and analysis of the relevant signal and background processes in proton-proton collisions at $\sqrt{s}=100$ TeV. The main results of this section are summarized in Table~\ref{tab:errors}, which contains quantitative estimates of the precision with which ingredients of the sum rules can be measured.

\subsection{Benchmark Points}
\label{sec:BPs}

Parameter spaces of both LH and MSCH models are constrained by the LHC searches for charge-2/3 top partners~\cite{Aaboud:2017qpr,Aaboud:2017zfn,Sirunyan:2017ynj,Aaboud:2018xuw,Sirunyan:2018omb,Aaboud:2018pii}. To satisfy these constraints, we choose benchmark points (BPs) in both models with top partner masses around $1.5$~TeV. The MSCH model also contains a charge-5/3 top partner, denoted as $x_5$ for mass eigenstate, which can produce a same-sign dilepton final state via the decay chain $x_5\rightarrow Wt\rightarrow \ell\nu\ell\nu b$. Stringent LHC limits~\cite{Sirunyan:2017jin,Aaboud:2018xpj} on this final state were taken into account in our choice of the BPs in this model. 

\begin{table}[t]
	\begin{center}
		\begin{tabular}{|c|c|c|c|c|c|c|c|}
		\hline
		Name & Mass [GeV] & Decays  & $\sigma(\T\bar{\T})$,~fb & $\sigma(qh\T)$,~fb & $\F(qh\T)
		$ & $\sigma(\T t hh)$,~fb & $\F(\T t hh)$ \\ \hline 
		\multirow{3}{*}{$\T$} & \multirow{3}{*}{1492} & Wb: 50\% & \multirow{3}{*}{$2250$} & \multirow{3}{*}{185} & \multirow{3}{*}{-13\%} & \multirow{3}{*}{14.4} & \multirow{3}{*}{-26\%} \\	
		& & tZ: 25\% & & & & & \\ & & tH: 25\% & & & & & \\ \hline
\end{tabular}
\caption{Top partner parameters at the Benchmark Point in the Little Higgs (LH) model.}\label{tab:BP_LH}
\end{center}
\end{table}

\begin{table}[t!]
	\begin{center}
	\begin{tabular}{|c|c|c|c|c|c|c|}
		\hline
		Name & Mass [GeV] & Decays  & $\sigma(qh\tp_i)$,~fb & $\F(qh\tp_i)$ & $\sigma(\tp_i t hh)$,~fb & $\F(\tp_i t hh)$ \\ \hline
		\multirow{3}{*}{$\tp_1$} & \multirow{3}{*}{ 1500 }  &  tH: 51.45\% & \multirow{3}{*}{$56.1$}  & \multirow{3}{*}{0} & \multirow{3}{*}{$28.0$}  & \multirow{3}{*}{0}\\ \cline{3-3}
& & Zt: 32.43\% & & & & \\ \cline{3-3}
& & WWt : 11.81\% & & & & \\ \hline
		\multirow{4}{*}{$\tp_2$} & \multirow{4}{*}{1632} &  tH: 37.51\% & \multirow{4}{*}{5.0}  & \multirow{4}{*}{-25\%} & \multirow{4}{*}{41.7} & \multirow{4}{*}{-45\%} \\ \cline{3-3}
		& & Wb: 32.84\% & & & &\\ \cline{3-3}
		& & Zt: 18.33\% & & & & \\ \cline{3-3}
		& & W$x_5$ : 6.08\% & & & & \\ \hline
		\multirow{4}{*}{$\tp_3$} & \multirow{4}{*}{1791}  & Zt: 42.92\%  
& \multirow{4}{*}{20.9} & \multirow{4}{*}{26\%} & \multirow{4}{*}{14.3} & \multirow{4}{*}{9.9\%}\\ \cline{3-3}
& & Wb: 26.06\% & & & & \\ \cline{3-3}
& & Z$\tp_2$: 12.76\% & & & & \\ \cline{3-3}
& & tH : 8.3\% & & & & \\		
\hline
	\end{tabular}
	\caption{Top partner parameters at the Benchmark Point in the Maximally Symmetric Composite Higgs (MSCH) model.}\label{tab:BP_MSH}
\end{center}
\end{table}

The Lagrangian parameters of the chosen BPs are as follows:

\paragraph{LH Model} $\lambda_1=1.48$, $\lambda_2=1.11$, and $f=811\GeV$.

\paragraph{MSCH Model} $\epsilon_{qQ}= 1.15$, $\epsilon_{tQ}=-1.40$, and $M=1500\GeV$.    \\

\noindent For the SM top, in both cases the benchmark parameters produce the Lagrangian mass $m_t\approx 150$~GeV, corresponding to the measured pole value of $172$ GeV. The masses and branching ratios of the top partners, along with cross sections of the relevant processes at the 100 TeV collider, are shown in Tables~\ref{tab:BP_LH} and~\ref{tab:BP_MSH}. To quantify the sensitivity of cross section measurements to the sign of $\Ytp$, we consider the {\it flip rate} $\F$, defined as
\beq
\F = 2 \, \frac{\sigma(\Ytp)-\sigma(-\Ytp)}{\sigma(\Ytp)+\sigma(-\Ytp)}.
\eeq{Fdef} 
The flip rate is the fractional change in the cross section when the sign of $\Ytp$ is flipped, while all other couplings and masses are unchanged. Flip rates for relevant reactions at our BPs are listed in Tables~\ref{tab:BP_LH} and~\ref{tab:BP_MSH}. Note that flip rates vanish for the lightest MSCH top partner $\tp_1$, since $Y_{\tp_1\tp_1h}=0$. For the other top partners, the flip rates can be used as a quick estimate of the accuracy of cross section measurement required to gain sensitivity to the sign of $\Ytp$. For the reactions we consider, cross sections are in the $10-100$ fb range and flip rates are in the $10-50$\% range, making it plausible that the sign of the coupling can be detected at a statistically significant level. The limiting factor will be separating the signals from SM backgrounds; this will be discussed in detail in the rest of this section.   

\subsection{Little Higgs}
\label{sec:LH}

In this subsection, we discuss how all ingredients of the naturalness sum rule in the Little Higgs model, Eq.~\leqn{LHsumrule}, can be measured experimentally at a 100 TeV collider. Here and in the rest of the paper, we assume integrated luminosity of 30 ab$^{-1}$~\cite{Arkani-Hamed:2015vfh,CEPC-SPPCStudyGroup:2015csa,Mangano:2017tke}. Monte Carlo (MC) event simulation for signal and background is performed within the \texttt{MadGraph5\_aMC@NLO} platform~\cite{Alwall:2014hca}. The {\tt Feynrules}~\cite{Alloul:2013bka} package was used to generate a UFO output model~\cite{Degrande:2011ua}. Events were then generated using {\tt MadGraph 5}~\cite{Alwall:2011uj}, the subsequent decays are handled by {\tt MadSpin}~\cite{Artoisenet:2012st}, posterior showering and hadronization are modeled by {\tt Pythia8}~\cite{Sjostrand:2007gs}, and finally detector effects are approximated by a parametrized detector simulation using {\tt Delphes3}~\cite{deFavereau:2013fsa}.

\subsubsection{Top Partner Mass and Branching Ratios}
\label{sec:LHmassBR}

The simplest process involving top partners at a hadron collider is pair-production, $pp\to T\bar{T}$. A very large sample of $T$ pairs will be collected at a 100 TeV collider, allowing for precise determination of the top partner mass and branching ratios. Both are important for testing the sum rules: $\mT$ enters the sum rule directly, while the branching ratios are needed to infer the coupling $\YT$ (see below).      

\begin{figure}[t!]
	\begin{center}
		\includegraphics[width=0.45
\linewidth]{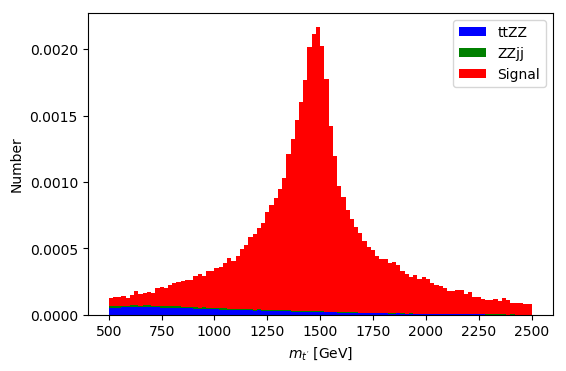}
		\includegraphics[width=0.45
	 \linewidth]{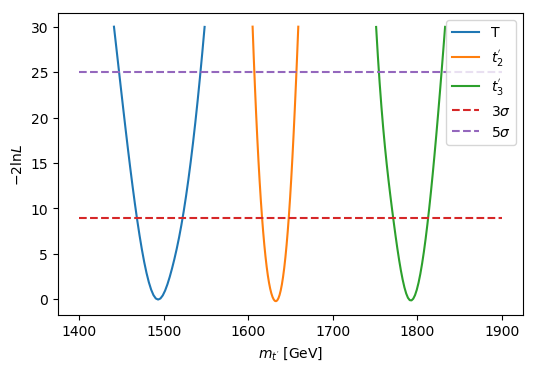}
	\end{center}
\vskip-0.5cm
	\caption{Left panel: Reconstructed invariant mass $m(\ell^+\ell^- j)$ for signal (LH model) and principal backgrounds, in the $4\ell+2j$ channel, after cuts described in the text. Right panel: Likelihood profile for the masses of top partners $T$ (LH model) and $\tp_2$, $\tp_3$ (MSCH model), measured using the fully reconstructed $Zt$ decay in the $4\ell+2j$ final state.}
	\label{fig:LH_mtp_distribution} 
\end{figure}

Top partner mass can be fully reconstruced in the decay $T \to Z t$, followed by a hadronic top decay. For simplicity, we consider the situation when this decay occurs on both sides of the event. To reduce QCD backgrounds, we assume that both $Z$'s decays leptonically. Since $\mT\gg m_t$, hadronic top decays would typically result in a single (fat) jet. We thus consider the final state with two lepton pairs, each with invariant mass close to $m_Z$, and two top-tagged fat jets. In the MC simulation, we apply the anti-$k_T$ reconstruction algorithm with cone size $R=1.0$, and require the final state to contain at least two jets with $p_T>400$ GeV. A Convolutional Neural Network (CNN) jet-tagger algorithm~\cite{Almeida:2015jua,deOliveira:2015xxd,Choi:2018dag}, which is described in detail in Appendix~\ref{App:JetTag}, is then applied, and the events with exactly two top-tagged jets are selected.  
Principal SM backgrounds to this final state, $ZZt\bar{t}$ and $ZZjj$ (with two QCD jets misidentified as top jets), have also been simulated. The distribution of events in the invariant mass $m(\ell^+\ell^- j)$ (including incorrect groupings dues to combinatorial uncertainties), as well as the corresponding likelihood profile for $\mT$, are shown in Fig.~\ref{fig:LH_mtp_distribution}. We conclude that $\mT$ can be measured in this channel with very small statistical error, approximately 1\%. In practice, systematic errors such as Jet Energy Scale (JES) uncertainty would likely dominate the measurement, probably giving an overall error of order a few~\%. This is sufficient for our purposes.     

\begin{table}[t!]
	\begin{center}
		\begin{tabular}{|l|c|c|c|c|c|c|c|} \hline
		 	Name &	Mass $\mtp$ & Br$(\tp\to Zt)$ & Br$(\tp\to th)$ & Br$(\tp\to Wb)$ & $g_{b\tp W}$ & $g_{t\tp h}$ & $\Ytp$ \\ \hline
		$T$ (LHT) &$1.2$ &$2.4$ &$2.3$ &$2.4$ &$1.2$ &$2.1$ &$2.9$	 \\ \hline	
		$\tp_2$ (MSCH) &$0.7$ &$4.4$ &$3.9$ &$8.6$ &$3.2$ &$5.7$ &$6.9$	 \\ \hline		
	    $\tp_3$ (MSCH) &$0.8$ &$0.8$ &$3.9$ &$1.3$ &$1.9$ &$2.8$ &$7.0$	 \\ \hline		
		\end{tabular}
		\caption{Estimated measurement precision, in \%, of the relevant parameters in the LH and MSCH models at a 100 TeV $pp$ collider with 30 ab$^{-1}$ data set. All errors are statistical only. See text for details. 
		}
	\label{tab:errors} 
	\end{center}
\end{table}

Since $T$ pair-production is dominated by strong interactions, the mass $\mT$ is the only non-SM parameter that enters $\sigma(T\bar{T})$ at leading order. Once this mass is measured, the cross section is completely fixed from theory, up to pdf uncertainties. Measurements of event rates in specific final states can then be used to deduce the branching ratios of various $T$ decay modes. For example, the event rate $N(ZtZt)$ in the fully-reconstructed final state discussed above is proportional to $[{\rm Br}(T\to Zt)]^2$. Using the $Z_{\ell\ell}+Z_{jj}+2j_t$ final state, we estimate that this rate can be measured with a statistical error of about $5$\%. The statistical effor of the branching ratio measurement can be estimated as 
\beq
\frac{\delta {\rm Br}(tZ)}{{\rm Br}(tZ)} = \frac{1}{2}\left(\frac{\delta N}{N}\oplus\frac{\partial \log \sigma}{\partial \log M_T}\frac{\delta M_T}{M_T}\right).  
\eeq{Br_error}
The derivative $\frac{\partial \log \sigma}{\partial \log M_T}$ was estimated numerically using {\tt MadGraph}. In this case, we find that the rate measurement dominates statistical uncertainty, resulting in a measurement of the branching ratio accurate to about $2.5$\%, see Table~\ref{tab:errors}. In reality, the measurement will also be affected by systematic errors, including pdf uncertainties, uncertainties in jet tagging probabilities, etc. While a quantitative analysis of these sources of error is beyond the scope of this work, we expect that these errors can be controlled at a few-\% level, sufficient for a highly non-trivial check of the sum rules.     

Similarly, the branching ratio ${\rm Br}(T\to th)$ can be inferred from the rate of $thth$ events. In this case, we consider the final state in which one of the Higgs bosons decays to $\gamma\gamma$, while the other Higgs decays to $b\bar{b}$. Each top is assumed to decay hadronically, typically producing a single fat jet. Likewise, the Higgs decaying to a $b\bar{b}$ pair is highly boosted in the lab frame, resulting in a single ``Higgs jet". We thus require 2 photons and 3 jets tagged as massive (non-QCD) jets in the final state. Principal SM backgrounds to this final state include $3j+\gamma\gamma$ (including both resonant diphotons from Higgs decays, and the non-resonant contriubution) and $t\bar{t}j+\gamma\gamma$. Imposing a cut $p_T>400$~GeV on all jets, we find a large event sample with $S/B\sim 1$, resulting in a measurement of the signal rate with a statistical error of about $5$\%. Finally, the channel $bWbW$ can be used to measure the remaining branching ratio, ${\rm Br}(T\to Wb)$. To suppress QCD backgrounds, we require that both $W$ bosons decay leptonically, resulting in a final state $\ell^+\ell^-+2b+\met$. The prinicpal backgrounds are from $t\bar{t}$ and $WWjj$ processes. We estimate that this signal rate can also be measured with a statistcial error of about 5\%. We conclude that with the assumed dataset, statistical errors on all three branching ratios will be about 2.5\%. While systematic errors, such as pdf uncertainties, are likely to somewhat degrade the precision, obtaining a few-\% measurement of the three branching ratios seems realistic.    

\subsubsection{Top Partner Yukawa Coupling: $pp\to q T h$} 

The second ingredient of the sum rule is the top partner Yukawa coupling $\YT$. We propose to infer this coupling, including the sign, from a measurement of the 
cross section $\sigma(pp\to q T h)$. Before demonstrating that this cross section can measured at the 100 TeV collider with sufficient precision, let us address the following important point. To infer the coupling $\YT$ from a measured cross section, one needs to have a theoretical prediction of this cross section in which $\YT$ is the {\it only} unknown parameter. In the LH model, the cross section depends on the top partner mass, $\mT$, and the couplings $g_{bT W}$ and $g_{tT h}$, in addition to $\YT$. (See the Feynman diagrams in Fig.~\ref{fig:FD_qhtp}.) The measurement of $\mT$ has already been discussed. The couplings $g_{bT W}$ and $g_{tT h}$ could potentially be inferred from the branching ratios ${\rm Br}(T\to Wb)$ and ${\rm Br}(T\to th)$ respectively. However, doing so requires an independent measurement of the total width of the top partner, $\Gamma_T$. Unfortunately, at our benchmark point, $\Gamma_T\approx 40$~GeV is too small to be directly measured: the width of the reconstructed top partner mass distribution in Fig.~\ref{fig:LH_mtp_distribution} is completely dominated by detector smearing effects. Instead, one can use a measurement of the single top-partner production cross section $\sigma(pp\to T q)$, which is proportional to $g_{bT W}^2$ and does not depend on any other non-SM parameters except $\mT$. To estimate the prosects for measuring this cross section, we simulated this process, with the decay channel $T\to Zt$ followed by a leptonic $Z$ decay and a hadronic top decay forming a top-jet. The dominant backgrounds, $Z+$jets and $t\bar{t}Z$, were also simulated. The selection criteria for this process are an opposite-sign lepton pair consistent with a $Z$ decay, and at least one top-tagged jet with $p_T>400$~GeV. We estimate a statistical error in the signal rate measurement of about 1\%. This signal rate can be converted into a cross section measurement by dividing it by the branching ratio ${\rm Br}(T\to Zt)$, which can be measured independently in the pair-production channel as discussed above. This method gives a statistical uncertainty on the measurement of $g_{bT W}$ of approximately $1.2$~\%, dominated by the branching fraction measurement. Once $g_{bT W}$ is known, the coupling $g_{tT h}$ can be inferred from the {\it ratio} of the branching ratios, $\frac{{\rm Br}(T\to th)}{{\rm Br}(T\to Wb)}$, which is independent of the total width. We estimate that this method will yield a measurement of $g_{tT h}$ with a statistical error of about 2\%.    

Now, let us turn to the main subject of this section, the measurement of $\sigma(pp\to q T h)$. This process can produce a number of signatures, depending on the decay chains of $T$ and $h$. Having studied a number of possible final states, we identified two channels where the signal can be separated from SM background and measured precisely.    

\subparagraph{$pp\rightarrow qhT\rightarrow (\gamma\gamma)+2j+X$ Channel.} In this channel, the Higgs boson decays to two photons. The top partner can decay in any of the available channels ($T\rightarrow Wb/Ht/Zt$), but both of its decay products are required to decay hadronically. Given that $M_T=1.5\TeV$ is well above the weak scale, the decay products of the $T$ are highly boosted. Each of their hadronic decays is typically reconstructed as a single jet, resulting in a final state with two high-$p_T$ jets. (The jet from t-channel production is typically not as hard as those from heavy $T$ decay, so we choose to require two hard jets in the final state.) The principal irreducible SM backgrounds in this channel include $hjj$, $tth$, and $\gamma\gamma jj$. We do not include reducible (instrumental) backgrounds in our study.           

\begin{table}
	\begin{center}
		\begin{tabular}{|c|c|c|}
			\hline
			Channel & Selection Cuts & Results  \\ \hline
			$(\gamma\gamma)+2j+X$ & $p_T(j^k)\geq 400$~GeV, $k=1,2$; & Table~\ref{tab:lh_qhtp} \\
			& $p_T(\gamma^1)\geq 100$~GeV;~~$p_T(\gamma^2)\geq 50$~GeV & \\
			& $m_{\gamma\gamma}\in [120, 130]$~GeV;~~$m_{jj}\in [1.0, 2.0]$~TeV & \\ \hline
			$(\ell^+\ell^-)+2j+X$ & {\bf Pre-cuts:} $p_T(j^k)\geq 400$~GeV, $k=1,2$; & Table~\ref{tab:zll_analysis} \\
			& $m_{\ell^+\ell^-} \in [80, 100]$~GeV;~~$p_T(Z)\geq 400$~GeV; & \\ 
			& 1 top-tag, 1 Higgs-tag & \\ 
			& {\bf Cut I:} $m(j_t \ell^+\ell^-)\in [1.0, 2.0]$~TeV & \\ \hline
			$(\gamma\gamma)+3j+X$ & $p_T(j^k)\geq 400$~GeV, $k=1\ldots 3$;~~$m_{\gamma\gamma}\in [115, 135]$~GeV;& Table~\ref{tab:lh_tpthh} \\
			& $\exists i, k\in \{1, 2, 3\}: m(j_i j_k) \in [1.0, 2.0]$~TeV & \\ \hline
		\end{tabular}
		\caption{Summary of selection cuts used for each of the three channels analyzed in this section. See text for more details.}
		\label{tab:CUTS}	
	\end{center}
\end{table}

\begin{table}
	\begin{center}
	\begin{tabular}{|c|c|c|c|c|c|c|}
		\hline
		Process & Pre-cuts & RF I &RF II &$\sigma$ &Sign $\sigma$  \\ \hline
		LH: $qhT\rightarrow (\gamma\gamma)+2j+...$ &$2.8\times10^3$ &$1540$ &744 &\multirow{4}{*}{$42$} &\multirow{4}{*}{$4.5$} \\ \cline{1-4}
		$hjj$ &$1.2\times10^4$ &$550$ &1766 & &  \\ \cline{1-4}
		$tth$ &$6.3\times10^3$ &$507$ &1273 & &  \\ \cline{1-4}
		$\gamma\gamma jj$&$4.8\times10^4$ &$344$ &2069  & & \\ \hline
	\end{tabular}
	\caption{First three columns: Signal and background event numbers in the $\gamma\gamma+2j+X$ channel in the LH benchmark model, in a 30 ab$^{-1}$ sample at a 100 TeV $pp$ collider. Fourth column: Poisson significance~\cite{Cowan:2010js} for discriminating signal+background model from pure background. Fifth column: Poisson significance for discriminating signal+background models with ``correct" and ``wrong" signs of $\YT$.}
	\label{tab:lh_qhtp}	
	\end{center}
\end{table}

In the Monte Carlo study of this signature, we apply anti-$k_T$ jet reconstruction algorithm with cone size $R=1.0$, and require the final state to contain at least two jets with $p_T>400$~GeV. We require at least two photons, with the leading photon $p_T>100$ GeV and the second-leading $p_T>50$ GeV. We further perform two ``pre-cuts", requiring the invariant mass of the two photons to be between 120 and 130 GeV, while the invariant mass of the two jets is between 1.0 and 2.0 TeV. The cuts for this and other analyses of this section are summarized in Table~\ref{tab:CUTS}. The event numbers in the signal and background samples that survive the cuts are listed in Table~\ref{tab:lh_qhtp}.   

\begin{figure}[t!]
	\begin{center}
		\includegraphics[width=0.75 \linewidth]{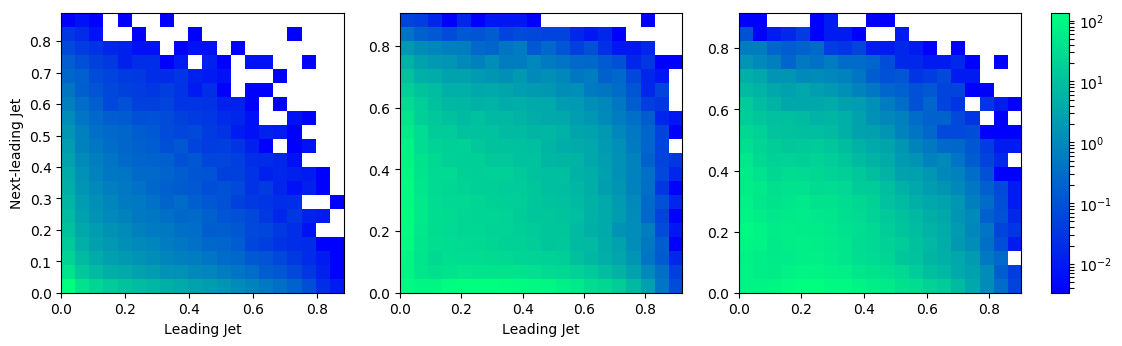}
		\includegraphics[width=0.5 \linewidth]{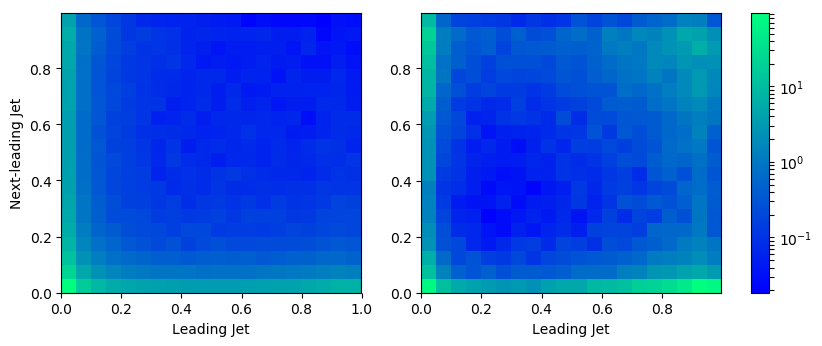}
	\end{center}
	\caption{Top: jet tagger-assigned probabilities of leading and subleading jet being a ``QCD jet", in the three samples (left to right: signal, $hjj$, and $\gamma\gamma jj$). Bottom: jet tagger-assigned probabilities of leading and subleading jet being a top jet, in the signal (left) and $tth$ (right) samples. Green/light pixels correspond to higher event rate, while blue/dark pixels correspond to lower event rate.}
	\label{fig:jet_probability} 
\end{figure}

To further separate signal and background, we performed an analysis based on Random Forest (RF) technique~\cite{Breiman2001}, implemented in {\tt scikit-learn}~\cite{scikit-learn,RandomForest}. RF is a machine-learning algorithm that aims to find a non-linear combination of input variables that is optimal for discriminating signal and background. The details of the RF analysis are presented in Appendix~\ref{App:RF}. The RF discriminator is provided with 19 input variables for each event. These include kinematic variables (such as $H_T$, $\met$, jet and photon $p_T$'s, etc.), as well as jet substructure information. For the latter, we use the CNN jet-tagger algorithm described in Appendix~\ref{App:JetTag}. The jet tagger uses the pattern of energy deposits in hadron calorimeter (HCAL) cells inside a jet to classify it according to its origin: light quark or gluon (``QCD jets"), bottom quark, $W/Z$, Higgs, or top. The output for each jet is a set of probabilities ${\cal P}_i$ for it to belong to each of the classes. In this analysis, we use the probabilities ${\cal P}_{\rm QCD}$ and ${\cal P}_t$, for the leading (in order of decreasing $p_T$) and first subleading jet. These are useful because the dominant backgrounds contain QCD or top jets, while signal events contain a $W$, $Z$ or $H$ jet (see Fig.~\ref{fig:jet_probability}). These four probabilities are provided as inputs to the RF discriminator. Combining kinematic and jet-substructure information, the RF algorithm outputs a single number optimized to distinguish between signal and background, $\xi_{\rm RF}$. The distribution of events in $\xi_{\rm RF}$ can then be fit by a sum of signal and background templates, providing a measurement of the signal rate. In this study, we have roughly approximated such a fit by simply counting signal and background events in two bins in $\xi_{\rm RF}$, with the highest $S/B$ ratios, and combining the statistical significance for each bin. The two bins are referred to as RF I and RF II in Table~\ref{tab:lh_qhtp}. For 30 ab$^{-1}$ integrated luminosity, the ``best" bin RF I contains about 1500 events and $S/B\sim 1$ (see Table~\ref{tab:lh_qhtp}). We conclude that the signal can be discovered in this channel with a very high statistical significance. Moreover, the measured signal rate in this channel can be used to extract $\YT$, including its sign. Flipping the sign of $\YT$ results in a 26\% shift in the signal cross section, see Table~\ref{tab:BP_LH}. Cut efficiencies are to a very good approximation unaffected by a change in the sign, so the rate of signal events passing the cuts is shifted by the same amount. With a 30 ab$^{-1}$ integrated luminosity, such a shift in the signal rate can be detected with a statistical significance of 4.5$\sigma$. Thus, if the LH model is correct, this channel will provide statistically significant evidence for the sign of $\YT$ consistent with the sum rule in Eq.~\leqn{LHsumrule}. It can also be used to extract the magnitude of $\YT$. Using {\tt MadGraph}, we numerically estimated the sensitivity of $\sigma(pp\to q T h)$ to $\YT$ as the coupling is varied around its the benchmark value: 
\beq
\frac{1}{\sigma}\frac{d\sigma}{d\YT} \approx -2.4
\eeq{sensitivity}
Taking also into account the sensitivity of the cross section to $g_{bT W}$ and $g_{tT h}$ and the uncertainties in these couplings discussed above, and combining all errors in quadruture, we estimate the statistical uncertainty of $\YT$ determination from this channel at about 3\%.

\subparagraph{$pp\rightarrow qhT\rightarrow (\ell\ell)+2j + X$   Channel} In this case, the top partner decays to $Zt$. The $Z$ then decays leptonically, while the top decays hadronically and produces a single top-jet. The Higgs boson is typically highly boosted in the lab frame, and we consider the decay to $b\bar{b}$, producing a single Higgs-jet in the final state. The major irreducible SM backgrounds are $t_qt_qZ_{\ell\ell}$ and $Z_{\ell\ell}jj$ with mis-identified jets.  

\begin{table}
	\centering
	\begin{tabular}{|c|c|c|c|c|c|c|}
		\hline
		Process & Pre-cuts & Cut I & RF I &RF II &$\sigma$ &Sign $\sigma$  \\ \hline
		LH: $qhT\rightarrow qh_{bb}Z_{\ell\ell}t_q$ &$2234$ &$1836$ &$1232$ &$366$ &\multirow{3}{*}{$ 52$} &\multirow{3}{*}{$4.2$} \\ \cline{1-5}
		$t_qt_qZ_{\ell\ell}$ & $1.4\times10^4$  &$5203$ &$64$  &$431$ & & \\ \cline{1-5}
		$Z_{\ell\ell}jj$&$5.6\times10^4$ &$1.9\times10^4$ &$211$ &$1431$ & & \\ \hline\hline
		MSCH: $qh\tp_3\rightarrow qh_{bb}Z_{\ell\ell}t_q$ &$586$ &440 &\multicolumn{2}{ c|}{263/70, 86/429} &23 &3.8 \\ \hline
	\end{tabular}
	\caption{First four columns: Signal and background event numbers in the $(\ell\ell)+2j + X$ channel in a 30 ab$^{-1}$ sample at a 100 TeV $pp$ collider. Fifth column: Poisson significance~\cite{Cowan:2010js} for discriminating signal+background model from pure background. Last column: Poisson significance for discriminating signal+background models with ``correct" and ``wrong" signs of $\YT$. For MSCH, we list the number of signal/total background events in the RF I and RF II signal bins.}
	\label{tab:zll_analysis}
\end{table}

In the Monte Carlo study of this signature, we again apply anti-$k_T$ jet reconstruction algorithm with cone size $R=1.0$, and require the final state to contain at least two jets with $p_T>400$~GeV. We further require two opposite-charge, same-flavor leptons, with invariant mass between 80 and 100 GeV consistent with $Z$ decay. We reconstruct the $Z$ transverse momentum $p_T^Z=p_T^{\ell_1}+ p_T^{\ell_2}$, and require $p_T^Z>400$~GeV. The jet tagger algorithm is then applied. We require that the event contain exactly one top-tagged and one Higgs-tagged jet. (We choose the tagger parameters corresponding to 50\% efficiency to ID the top and the Higgs jet, respectively. For mistag rates and further details, see Appendix~\ref{App:JetTag}.) The event numbers passing these pre-cuts in a 30 ab$^{-1}$ data sample are listed in Table~\ref{tab:zll_analysis}. The invariant mass of the jet identified as top and the two leptons from $Z$ decay is expected to match the top partner mass in the signal, and we require that this invariant mass be between 1 and 2 TeV (``Cut I" in Table~\ref{tab:zll_analysis}). At this point, $S/B\sim 1/10$ is achieved. 

To further discriminate between signal and background, we once again construct an RF discriminator. In this case, the RF inputs do not include jet substructure information, since it has already been used. The best bin in RF discriminator output $\xi_{\rm RF}$, labeled RF I in Table~\ref{tab:zll_analysis}, contains about 1200 signal events for a 30 ab$^{-1}$ data sample, with excellent $S/B\sim 5$. The measurement of the signal rate in this channel on its own can discriminate between the ``correct" (as required by the sum rule) and ``wrong" sign of $\YT$ with a statistical significance of about 4$\sigma$. Combining the two channels considered in this subsection provides a total significance of over 6$\sigma$.           
    
\subsubsection{Top Partner Yukawa Coupling: $pp\to T t h h \to (\gamma\gamma)+3j+X$}    
    
The second interesting channel sensitive to the sign of $\YT$ is $pp\rightarrow Tthh$, see Fig.~\ref{fig:FD_tpthh}. We considered the final state in which one of the Higgs bosons decays to a photon pair, while all other decays are hadronic. Since the top and at least one of the Higgs bosons in the final state predominatly come from $T$ decays, they tend to be highly boosted in the lab frame, and their decay products are reconstructed as a single massive jet. Likewise, hadronic decays of the second $T$ are typically reconstructed as a pair of massive jets (in the $tZ$ and $th$ decay channels) or a massive jet and a light jet (in the $bW$ channel). Therefore we require that the final state contain 2 photons and at least three jets, tagged as massive (non-QCD) jets by the jet tagger. (In this case, we do not use the jet tagger to discriminate among various types of massive jets such as Higgs, top, etc.) Further, we impose the following pre-cuts:

\begin{itemize}
	\item $p_T\geq 400$~GeV for each of the three massive jets;
	\item Two-photon invariant mass consistent with Higgs decay: $m_{\gamma\gamma}\in [115, 135]$~GeV;
	\item At least one out of three pair-wise combinations of the three massive jets has invariant mass roughly consistent with $T$ decay: $m_{j_ij_k}\in [1000, 2000]$~GeV.  
\end{itemize}        

\begin{table}[t]
	\begin{center}
	\begin{tabular}{|c|c|c|c|c|c|c|}
		\hline
		Process & Pre-cuts  & RF  & $\sigma$ &Sign $\sigma$  \\ \hline
		LH: $Tthh\rightarrow (\gamma\gamma)+jjj+...$ &$148$ &$126$  & \multirow{3}{*}{ $14$} &\multirow{3}{*}{$2.4$}  \\ \cline{1-3}
		$ttjh$ &$183$ &$48$&  &\\ \cline{1-3}
		$ttVh$&$5.6$ &$1.6$& &  \\ \hline\hline
		MSCH: $\tp_2thh\rightarrow (\gamma\gamma)+jjj+...$ &$487$ &$467/83$ & $34$ &$8.5$  \\ \hline
	\end{tabular}
	\caption{First two columns: Signal and background event numbers in the $(\gamma\gamma)+3j+X$ channel in a 30 ab$^{-1}$ sample at a 100 TeV $pp$ collider. Third column: Poisson significance~\cite{Cowan:2010js} for discriminating signal+background model from pure background. Last column: Poisson significance for discriminating signal+background models with ``correct" and ``wrong" signs of $\YT$. For MSCH, we list the number of signal/total background events in the RF signal bin.}
	\label{tab:lh_tpthh}
	\end{center}
\end{table}

With these pre-cuts, we find that the most significant surviving SM backgrounds are $t\bar{t}jh$ and $t\bar{t}Vh$ where $V=W/Z$, decaying hadronically. For the LH benchmark point parameters, $S/B\sim 1$ after the pre-cuts (see Table~\ref{tab:lh_tpthh}). To further improve background rejection, we again apply the Random Forest algorithm; see Appendix~\ref{App:RF} for details. The resulting statistical significance of the signal observation in this channel is about 14$\sigma$, while the LH model with flipped sign of $\YT$ can be distinguished from the ``correct" model at the $2.4\sigma$ level. While not as significant as the $qhT$ channel, this channel can provide valuable independent confirmation of the sign of $\YT$. This channel will have a more important role to play in the MSCH model.  
    
\subsubsection{Summary: Testing the Sum Rule in LH Model}

The analysis of this section established that at the chosen benchmark point in the LH model, the top partner mass and Yukawa coupling will be measured with statistical uncertainty of approximately 1\% and 3\%, respectively. Including systematic errors will likely somewhat degrade the precision, but it seems very likely that a few-\% measurement of these quantities will be achieved. Of course, the SM top parameters will be by then known with much better precision. We conclude that the 100 TeV collider program with 30 ab$^{-1}$ integrated luminosity will provide a quantitative test of the sum rule, Eq.~\leqn{LHsumrule}, including the relative signs of the two terms, at a few-\% level.      
    
\subsection{Maximally Symmetric Composite Higgs}

In the MSCH model, both quadratic and logarithmic divergences are canceled, and there are two sum rules, Eqs.~\leqn{MSCH_sumrules}. There are three top partner mass eigenstates, $\tp_1$, $\tp_2$ and $\tp_3$. The masses and Yukawa couplings of these states need to be measured to test the sum rules. Since the $\tp_1$ has vanishing Yukawa coupling, it does not contribute to the sum rules. In this section, we will focus on the phenomenology of $\tp_2$ and $\tp_3$. The set of measurements that need to be performed for each of these states is essentially the same as for the top partner in the LH model discussed above. The only new feature is that the contributions of each of the top partners to the rates need to be disentangled to extract their parameters individually. 

To assess the feasibility of such measurements at the 100 TeV collider, we performed a detailed Monte Carlo study, parallel to the case of LH model described above. The simulations are based on our implementation of the MSCH model in {\tt FeynRules}. (The model files are publicly available~\cite{Baidu}.) In this section we breifly summarize the results of this study.     

The mass of both $\tp_2$ and $\tp_3$ can be fully reconstructed in the $Zt$ decay channel, with a leptonic $Z$ and a hadronic top forming a single top-jet. This reconstruction was studied in the MC simulation, with the signal given by the pair-production and decay of $\tp_2$ and $\tp_3$ in the benchmark MSCH model, and the rest of the analysis identical to the LH case discussed in Sec.~\ref{sec:LHmassBR}. Effects of detector resolution are included in the analysis, and dominate the widths of the reconstructed resonances. The widths are sufficiently small for the two resonances to be clearly separated. The likelihood profiles for the masses of $\tp_2$ and $\tp_3$ are shown in the right panel of Fig.~\ref{fig:LH_mtp_distribution}. Statistical precision of the measurement of each mass is estimated to be close to 1\%, see Table~\ref{tab:errors}.

The next step is to measure branching ratios. The procedure is the same as for the LH top partner, and the only new issue is how to separate the two states. Branching ratio measurements in the $tZ$ and $th$ channels rely on final states where the decay is fully reconstructed, and therefore events with $\tp_2$ and $\tp_3$ production can be easily distinguished from each other by the reconstructed mass. To measure the $bW$ branching ratio, the LH analysis used the $2(Wb)$ final state with two leptonic $W$'s, making mass reconstruction impossible. To avoid this problem, in the MSCH case we used a different final state, $Wb+Zt$, in which the two top partners can be distinguished by their mass. The projected statistical uncertainties on the branching ratio measurements are summarized in Table~\ref{tab:errors}.  

The final step is to measure the cross sections of $pp\rightarrow qh\tp$ and $pp\to T t h h$ processes, and to extract $\Ytp$. Again, the analysis of the LH model applies here as well, with the only new issue being the simultaneous presence of multiple top partners. In all final states considered in the LH analysis above, the top partner decay is fully reconstructed, either as a pair of jets, or as a leptonic $Z$ in association with a (massive) jet. It should be straighforward to distinguish the events involving various top partners on an event-by-event basis. Moreover, the three top partner masses would be measured independently and with high precision using the pair-production process, and this information can be used in these channels, {\it e.g.} to reduce combinatoric errors in mass reconstruction. For the estimates in Table~\ref{tab:errors}, we assume perfect separation of the events involving different top partners, and analyze each one independently, using the procedure outlined in Section~\ref{sec:LH}. We find that the best measurement of the Yukawa coupling is provided by different channels: in the case of $\tp_2$, the best channel is $\tp_2 thh \to (\gamma\gamma)+3j+X$, see Table~\ref{tab:lh_tpthh}; while for the case of $\tp_3$, the best channel is $qh \tp_3 \to qhZt$, see Table~\ref{tab:zll_analysis}. The reason can be seen from the production cross sections and branching ratios in Table~\ref{tab:BP_MSH}. The resulting statistical error in the determination of the top partner Yukawa is about 7\% for both $\tp_2$ and $\tp_3$. This error includes the propagated error from mass and branching ratio measurements. We expect that systematic errors can be controlled at a similar or lower level, so that the total uncertainty on the Yukawas would be at most of order 10\%. Given that the top partner mass measurements will have much smaller uncertainty, we conclude that the set of measurements outlined here will result in a quantitative test of the sum rules, including signs, at a 10\% level.   

\section{Discussion}
\label{sec:discussion}

Naturalness of EWSB has been the main theoretical motivation for new particles and interactions appearing around the TeV scale. The main role of such new particles is to cancel the large quantum corrections to the Higgs potential and render it UV insensitive. These cancelations present the essence of the new physics models, and will manifest themselves in naturalness sum rules: relations among the couplings and masses of the SM particles and their new particle partners. We have derived these sum rules for the first time and presented them  both for same spin and opposite spin partners. We find a separate sum rule for the cancelation of the quadratic and logarithmic divergences. Depending on the particular model one or both may hold, and usually the sum rules are separately satisfied in the gauge and the fermionic sectors for same spin partners. We presented both the very general form of the sum rules as well explicit demonstration of how these sum rules are obeyed in two benchmark models: the Little Higgs and the Maximally Symmetric Composite Higgs. The LH model satisfies only the first sum rule for the top sector, ensuring the cancelation of the quadratic divergences, while  in the MSCH both  sum rules for the quadratic and logarithmic divergences hold.

If and when new particles were discovered, testing whether their masses and couplings satisfy our sum rules would provide the most direct test of naturalness. Thus it is essential to establish effective strategies to experimentally test the sum rules at future colliders. In the second half of this paper we have outlined how this goal may be achieved at the proposed 100 TeV $pp$ collider, using fermionic (spin-1/2) top partners in LH and MSCH models as a concrete example. We focused on measurements that are sensitive to the sign of the top partner coupling to the Higgs boson, since determining this sign is crucial for testing the sum rule. We performed a detailed Monte Carlo study of the relevant collider phenomenology, including the dominant SM backgrounds. Using traditional rectangular cuts as well as more advanced machine learning techniques, we constructed analyses that are capable of measuring top partner masses and their Yukawa couplings to the Higgs with excellent precision. Our findings can be summarized as it follows:

\subparagraph{Little Higgs model:} Given the chosen benchmark point, the mass of the top partner can be measured with statistical uncertainty of 1\%, while the top partner Yukawa can be determined with a 3\% statistical uncertainty.  
\begin{itemize}
\item In the $pp\rightarrow qhT\rightarrow (\gamma\gamma)+2j+X$ channel our analysis reaches a very high ($>5 \sigma$) significance for top partner discovery, and the statistical significance of 4.5$\sigma$ for the Yukawa sign determination. 
\item In the $pp\rightarrow qhT\rightarrow (\ell\ell)+2j + X$ channel the analysis achieved a high ($>5 \sigma$) significance for discovery, and a 4.2$\sigma$ significance for Yukawa sign determination. 
\item In the $pp\to T t h h \to (\gamma\gamma)+3j+X$ channel we achieve a high ($>5 \sigma$) significance for discovery, and a 2.4$\sigma$ significance for Yukawa sign determination.
\end{itemize}

\subparagraph{Maximally Symmetric Composite Higgs model:} In the MSCH model there are three top partner mass eigenstates, $t'_{1},t'_{2}$ and $t'_{3}$. The lightest mass eigenstate $t'_{1}$ has a vaniching Yukawa, $Y_{t'_{1}t'_{1}h}=0$, and does not contribute to the sum rules, so we concentrate our analysis on $t'_{2}$ and $t'_{3}$. We showed that the mass of each top partner can be reconstructed using the $Zt$ decay channel. The widths are small enough for the two resonances to be clearly separated. The statistical uncertainty for the measurement of the top partner masses is about 1\%, while their Yukawa couplings $\Ytp$ can be determined with a 7\% statistical uncertainty. 
\begin{itemize}
\item In the $pp\rightarrow qh\tp_3\rightarrow (\ell\ell)+2j + X$ channel the analysis achieves a high ($>5 \sigma$) significance for discovery of $\tp_3$, and a 3.8$\sigma$ statistical significance for $\tp_3$ Yukawa sign determination. 
\item In the $pp\to \tp_2 t h h \to (\gamma\gamma)+3j+X$ channel, we achieve a high ($>5 \sigma$) statistical significance for discovery of $\tp_2$, as well as for the deterimination of the sign of its Yukawa coupling.
\end{itemize}

Taken together, our results indicate that, if one of the new physics models studied here is realized in nature, the 100 TeV collider will be able to provide a highly non-trivial quantitative test of the naturalness sum rules. Thus, the sum rules derived in this paper are not only theoretically interesting as a general, elegant consequence of naturalness of electroweak symmetry breaking, but may also provide a key experimental test of this concept.

\section*{Acknowledgements}
The authors are grateful for conversations with Yongchen Wu for the collider part. C.C. thanks the Aspen Center for Physics (supported in part by the NSF grant PHY-1607611) and the ITP of the CAS in Beijing for their hospitality while working on this project. T.M. thanks the Cornell Particle Theory group for its hospitality while finishing this project.  C.C., L.H. and M.P. are supported in part by the US National Science Foundation grant PHY-1719877. C.C. is supported in part by the BSF grant 2016153. J.S. is supported by the NSFC under grant No.11647601, No.11690022, No.11675243 and No.11761141011 and also supported by the Strategic Priority Research Program of the Chinese Academy of Sciences under grant No.XDB21010200 and No.XDB23000000.  T.M. is supported in part by project Y6Y2581B11 supported by 2016 National Postdoctoral Program for Innovative Talents. F.F.F is supported in part by project Y8Y2411B11, China Postdoctoral Science Fundation. L.H. is supported by China Scholarships Council ( No. 201704910746 ).

\begin{appendix}

\section*{Appendix}

\section{Derivation of the Gauge Sector Sum Rules}
\label{App:gaugesumrules}

 The mass matrices for the vector bosons can also be parametrized as 
\bea
M_V =M_{V0} +f U_V(\frac{H}{f}),
\eea
where the $M_{V0}$ and $U_V(\frac{H}{f})$ are contributions to the mass matrix that are independent/dependent the on Higgs field.     
The quadratically divergent contribution to the Coleman-Weinberg one loop effective potential from gauge bosons is proportional to  
\bea
V_g(H) \sim  \mbox{Tr}[M_V^2] \Lambda^2.
\eea     
Similar to the top sector, we can get the main condition for the cancelation of the quadratic divergences from vector bosons loop  to be
\bea \label{eq:naturalness2}
\frac{\partial^2 \mbox{Tr}[M_V^2]   }{\partial H^2} \vert_{H=0} =0.
\eea

Following the analysis we performed in the top sector, we can again find that the Higgs $h$ tadpole in the $V_g$ after EWSB is proportional to (\ref{eq:naturalness2}) which is driving the $H$ 
quadratically divergent Higgs mass term before EWSB. If the gauge contributions to the  Higgs mass are not quadratically divergent, a sum rule for the terms linear in $h$ in $V_g$ must be satisfied up to $\mathcal{O}(\tilde{v}^2/f^2)$
\bea
\frac{\partial \mbox{Tr}[M_V^2]}{\partial h} \vert_{h=0} =0 +\mathcal{O}(\tilde{v}^2/f^2).
\eea    
We know the term $\partial M_V^2 /\partial h  \vert_{h=0}$ is the Higgs  linear coupling matrix to vector bosons pairs in EW eigenstate. We can rotate the vector bosons to mass eigenstates and the Higgs linear coupling matrix in mass eigensates as
\bea
g_{VVh} \equiv  L_g  \frac{\partial M_V^2}{\partial h} \vert_{h=0} L_g^\dagger,
\eea     
where $L_g$ is the rotation matrix that diagonalizes the vector boson  masses. Thus we obtain  a sum rule for the linear couplings of the Higgs to the gauge bosons:
\bea
\mbox{Tr}[g_{VVh}] =\mbox{Tr}[L_g  \frac{\partial M_V^2}{\partial h} \vert_{h=0} L_g^\dagger ] = \mbox{Tr}[\frac{\partial M_V^2}{\partial h} \vert_{h=0} ] =0 +\mathcal{O}(\tilde{v}^2/f^2). 
\eea    

The sum rule for the vanishing of the log  divergent contributions  from vector boson loop can be obtained in the same way. The condition for the vanishing of the log divergences is 
\bea
\frac{\partial^2}{\partial H^2} \mbox{Tr}[(M_V^2)^2]\vert_{H=0} =0+ \mathcal{O}(\tilde{v}^2/f^2). 
\eea

The Higgs tadpole after EWSB is again proportional to this term so the sum rule will be
\bea
\frac{\partial \mbox{Tr}[M_V^4]}{\partial h} \vert_{h=0} =  0+ \mathcal{O}(\tilde{v}^2/f^2).  
\eea
Rotating into the mass eigenstates, we can get the sum rule for physical sates 
\bea
\mbox{Tr}[g_{VVh}  M_V^2]= \mbox{Tr}[(L_g \frac{\partial M_V^2}{\partial h} \vert_{h=0} L_g^\dagger) (L_g M_V^2\vert_{h=0} L_g^\dagger)]   =\frac{\partial \mbox{Tr}[M_V^4]}{\partial h} \vert_{h=0}=0+ \mathcal{O}(\tilde{v}^2/f^2), 
\eea  
where $M_V$ is the mass matrix for physical vector bosons very similar to that for the top sector.

\section{Derivation of Sum Rules with Opposite Spin Partners} 
\label{App:supersumrules}

In this Appendix we present the derivation of the sum rules if the cancellation is due to opposite spin partners like in SUSY models. Since SUSY models usually contain two Higgs doublets we will also assume this but the final sum rule applies to models with arbitrary number of Higgs doublets.
For deriving our sum rule we will be treating  the CP even neutral component in each Higgs doublet as a background field. The quadratic divergences in the Coleman Weinberg potential  from loops of  spin $j$ particles  can be parametrized as 
\bea \label{eq:susy}
V_{SUSY}(H_1, H_2) &=&  \text{Str}[ M^2(H_1, H_2) ] \Lambda^2=\sum_j (-1)^{2j +1} (2j+1) \text{Tr}[M_j^2(H_1, H_2) ] \Lambda^2 \nonumber \\
&=&\text{Tr}[M_S^2 ] -2 \text{Tr}[M_F M_F^\dagger] +3   \text{Tr}[M_V^2],   
\eea 
where $M_S^2$ is the scalar mass matrix. 
If the mass matrix of the two Higgs doublets contains no quadratic divergence we must have
\bea
\frac{\partial^2 V_{SUSY}(H_1, H_2)}{\partial H_i \partial H_j}|_{H_1 =H_2 =0} =0. 
\eea   
As usual we assume that the two neutral Higgs bosons acquire VEVs 
\bea
 H_i  =v_i +h_i. 
\eea
Following our previous line of argument, the tadpole of the physical Higgs boson(s) must vanish, leading to the expression
\bea
&& \frac{\partial V_{SUSY}(H_1, H_2)}{\partial h} |_{h =0}  \nonumber \\
&=& \frac{\partial H_i }{\partial h}  |_{h =0} \frac{\partial V_{SUSY}(H_1, H_2)}{\partial H_i} |_{H_1 =H_2=0} +\frac{\partial H_i H_j }{\partial h}  |_{h =0} \frac{\partial^2 V_{SUSY}(H_1, H_2)}{2\partial H_i  \partial H_j } |_{H_1 =H_2=0} \nonumber  \\
&=&  \frac{\partial H_i }{\partial h}  |_{h =0} \frac{\partial V_{SUSY}(H_1, H_2)}{\partial H_i} |_{H_1 =H_2=0} =0
\eea  
Within a renormalizable SUSY theory the mass matrices depend only quadratically on the Higgs field and the above relation should be exact. In more general non-renormalizable theories we would get corrections of order $(v/\Lambda )^2$ where $\Lambda$ is the scale suppressing non-renormalizable operators.  We can apply the vanishing of the tadpole to the  general expression of the potential in Eq.~(\ref{eq:susy})  and  rotate the weak eigenstates  to mass eigenstates, yielding  the general sum rule for opposite spin partners 
\bea \label{eq:SUSY_sumrule_1}
\text{Tr}[g_{SSh}] -2\text{Tr}[Y_M M_D^\dagger + M_D Y_M^\dagger ] +3  \text{Tr}[ g_{VVh} ]=0,  
\eea 
 where $g_{SSh} \equiv  L_S  \frac{\partial M_S^2}{\partial h} \vert_{h=0} L_S^\dagger$ is the physical Higgs trilinear coupling matrix to a pair of scalars ($L_S$ is the rotation matrix diagonalizing the mass matrix for the scalars). If CP is conserved, we can further simplify the sum rule to
 \bea \label{eq:SUSY_sumrule_2}
 \text{Tr}[g_{SSh}] -4\text{Tr}[Y_M M_D] +3  \text{Tr}[ g_{VVh} ]=0.
 \eea 
We can specify the form of this sum rule to the top and the gauge sectors assuming that the divergences cancel independently in the two sectors (as they do in SUSY models).

\subsection{Top/stop sector}
We assume that the top partners are scalars (stops) or vectors.  The sum rule in terms of  the physical couplings of the top and its partners to the Higgs is 
 \bea \label{eq:SUSY_sumrule_3}
 \sum_i g_{\tilde{t}_i  \tilde{t}_i h}  -4y_{t} m_t +3 \sum_j g_{\tilde{V}_j \tilde{V}_j h}  =0,
 \eea  
 where $g_{\tilde{t}_i  \tilde{t}_i h} $ and $g_{\tilde{V}_j \tilde{V}_j h}$ are the Higgs trilinear coupling to stops   $\tilde{t}_i$ and vector top   partners $\tilde{V}_j$.          
There would be a similar sum rule in other fermionic sectors. 

 \subsection{Gauge/Gaugino/Higgsino sector} 

We assume that the partners of the gauge bosons and the Higgs are fermions (gauginos and Higgsinos) and they can mix with each other after EWSB.  The sume rule including the effects of the neutralinos and charginos is
\bea \label{eq:SUSY_sumrule_4}
&& \text{Tr}[g_{S^+S^-h}] -4\text{Tr}[Y_{C^+ C^- h } M_C^{\dagger} ] +3  \text{Tr}[ g_{W^+W^-h} ] + \nonumber \\
&&  \text{Tr}[g_{S^0S^0h}] -4\text{Tr}[Y_{NNh } M_N^{\dagger} ] +3  \text{Tr}[ g_{ZZ h} ]=0,
\eea 
 where $g_{S^+S^- h}$ and $g_{S^0S^0h}$ are the Higgs trilinear coupling matrices to electrically charged and neutral scalars in mass eigenstate, $Y_{C^+C^- h }$ and $Y_{NNh}$  are the Higgs trilinear coupling matrices to  charged and neutral fermions in mass eigenstates, $M_C$ and $M_N$ are the diagonal mass matrices of the physical  charginos and neutralinos and $g_{W^+W^-h}$ and  $g_{ZZ h} $ are the Higgs trilinear couplings to the SM $W^{\pm}$ and $Z$ bosons.        
If there is a separate cancelation in the electrically charged and neutral sectors the two lines in (\ref{eq:SUSY_sumrule_4}) will separately vanish.  Note that the $g_{S^0S^0h}$ term includes the Higgs self-coupling.


  

\section{Jet Tagging}
\label{App:JetTag}

Jet tagging plays a very important role in the collider analysis of Section 4. In this Appendix, we describe the jet-tagging algorithm that was constructed and used in that analysis. The algorithm is based on a Convolutional Neural Network (CNN), which combines calorimeter and tracker information for each jet to assign  probabilities that the jet is due to a top, Higgs, or $W/Z$ decay, a $b$ quark, or is a ``light" jet (originated by a $u/d/s/c$ quark or a gluon).       

\begin{table}[t]
	\begin{center}
		\begin{tabular}{|c|c|c|c|c|c|c|c|}
			\hline
			$p^{\mathrm{fat jet}}_{T}$  &j  &b  &t & Z & W & H\\ \hline
			200-400 GeV      &100\% &100\% &84.38\% &91.10\%  &90.93\%  &72.07\% \\ \hline
			400-800 GeV     &100\% &100\% &92.27\% &94.37\%  &94.06\%  &80.68\%  \\ \hline
			$\geq$ 800 GeV &100\% &100\%   &94.01\% &95.42\%   &95.04\%   &81.18\% \\ \hline
		\end{tabular}
		\caption{Jet-finding efficiencies as a function of jet $p_T$ in the 6 SM samples used for training and validating the CNN tagger.}
		\label{tab:jet_efficiency}
	\end{center}
\end{table}

\begin{figure}[t!]
	\begin{center}
		\includegraphics[width=0.45 \linewidth]{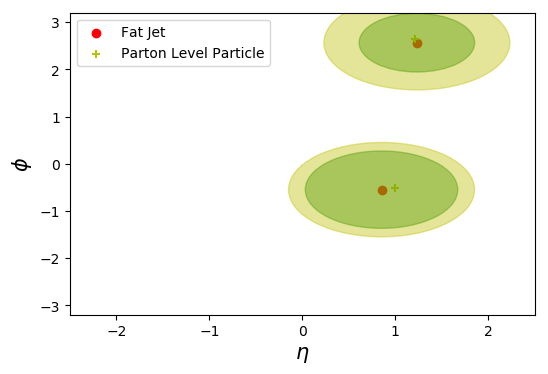}
		\includegraphics[width=0.45 \linewidth]{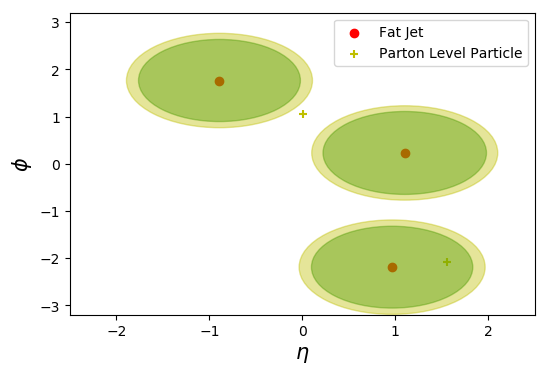}
	\end{center}
	\caption{$pp\rightarrow t\bar{t}$ final states. The red dot is the center of the identified jet, the yellow cross is the parton-level top, the yellow region is the clustering cone $R=1.0$, and the green region is the actual extent of the jet. In the left panel two top jets are identified, while in the right panel only one top jet identified.}
	\label{fig:pptt} 
\end{figure}

To train and test the jet tagger algorithm, we generated large event samples for SM pair-production processes, $pp\rightarrow XX$ with $X=j, b, t, h, W, Z$. The samples were generated by {\tt MadGraph} (parton-level production and final-state decays), {\tt Pythia} (showering and hadronization), and {\tt Delphes} (fast detector simulation). In the case of $t/h/W/Z$ final states, fully hadronic decays were simulated, and lower $p_T$ throesholds were chosen sufficiently high so that the decay products are typically collimated. Anti-kT jet algorithm~\cite{Cacciari:2008gp}, with jet clustering radius $R=1.0$, is applied to these samples. We used the {\tt FastJet}~\cite{Cacciari:2011ma} implementation of the anti-kT algorithm. Jet-finding efficiencies for each event class are shown in~\tab{jet_efficiency}. Two representative events from the $t\bar{t}$ sample are shown in Fig.~\ref{fig:pptt}. 

To classify the jets, each of them is converted into a ``tensor image", using the following procedure:

\begin{itemize}
	\item Choose a coordinate system centered at the center of the jet, so that $\left(\eta=0, \phi=0\right)$.
	\item Consider a square region in the $(\eta, \phi)$ plane, of size $1.0\times 1.0$, centered at the origin. By construction, the clustered jet is fully (or almost fully) contained within this region. Divide this image into $50\times50$ equal-sized pixels. (The angular size of each pixel roughly corresponds to the resolution expected for calorimeters at a future hadron collider.) 
	\item For each pixel, record the total $p_{T}$ of particles that fall within that region, as well as and the multiplicity, for both Track and Tower classes from {\tt Delphes}, which corresponds to the number of charged and neutral particles detected by the calorimeter.
	\item Combine this information in a tensor with dimensions $50\times 50\times 4$.
\end{itemize}

The tensor images of jets serve as inputs to the CNN constructed using the pyTorch~\cite{paszke2017automatic} enviroment. The CNN consists of the following elements:

\begin{itemize}
	\item 4 convolutional layers with a ReLU acctivation function;
	\item Two Maxpooling layers;
	\item Classification block layers, including two linear layers with Dropout of 50\% and ReLu activation function. The final linear layer classifies the jet images into 6 categories: top fat jet, Higgs fat jet, W fat jet, Z fat jet, b fat jet and light jet.
\end{itemize} 

Further, each jet is assigned randomly to either the training sample or the validation sample. In each sample, jets are divided into three bins according to their $p_{T}$: $200\GeV<p^{\mathrm{jet}}_{T}<400\GeV$, $400\GeV<p^{\mathrm{jet}}_{T}<800\GeV$ and $p^{\mathrm{jet}}_{T}>800\GeV$. Jets with $p^{\mathrm{jet}}_{T}<200\GeV$ are discarded, since they are not expected to exhibit relevant sub-structure. The CNN is trained using the traning sample, separately for each $p_T$ bin. The performance is then tested using jets in the validation sample. The tagger performance can be characterized by the Receiver Operating Characteristic curve (ROC curve). For each pair of jet classes $a$ and $b$, the ROC curve shows ``tag efficiency" (probability of correctly tagging the jet of class $a$ as $a$) on the horizontal axis, and $1-$ mistag rate (the probability of incorrectly tagging jet of class $b$ as $a$) on the vertical axis.  Benchmark working points used in the collider analysis correspond to $50\%$ and $80\%$ efficiency for the relevant jet class ($t$, $H$, or $W/Z$, depending on the analysis). These benchmark points are labeled on each of the curves, and the corresponding mistag rates are listed in the plot legend. 

\begin{figure}[t!]
	\begin{center}
		\includegraphics[width=0.45 \linewidth]{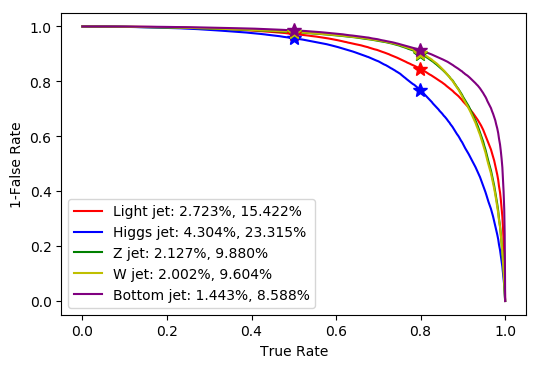}
		\includegraphics[width=0.45 \linewidth]{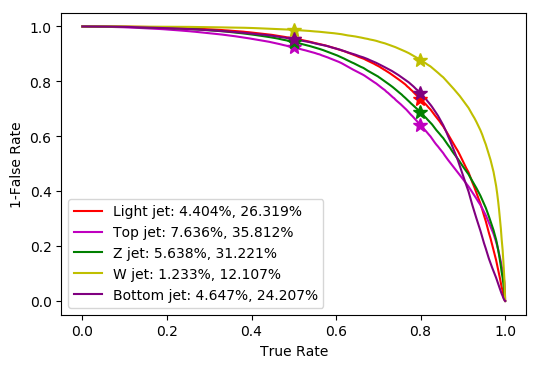}
		\includegraphics[width=0.45 \linewidth]{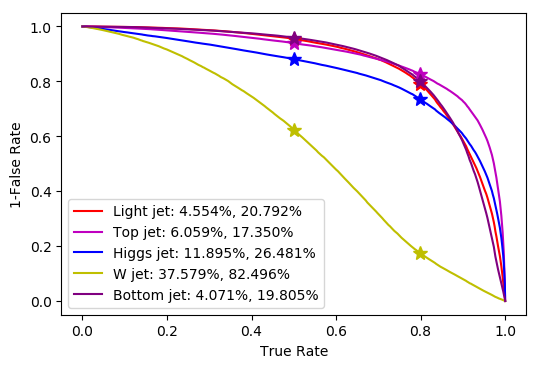}
		\includegraphics[width=0.45 \linewidth]{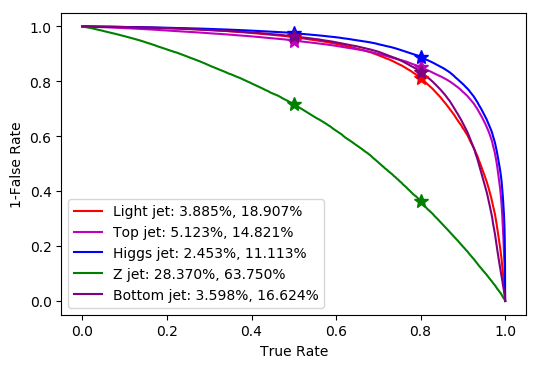}
	\end{center}
	\caption{Jet tagging performance for the first region selection $400 < p^{fat jet}_{T}  < 800\GeV$ by ROC curve. From the top left plot, we have the jat tagging efficiency for top, Higgs (top right), Z (bottom left) and W (bottom right) jet respectively. The y-axis are the 1 minus false rate. In the plot labels we display the miss identification for 50\% and 80\% benchmark points.}
	\label{fig:jet_tagger_400}
\end{figure}

\begin{figure}[t!]
	\begin{center}
		\includegraphics[width=0.45 \linewidth]{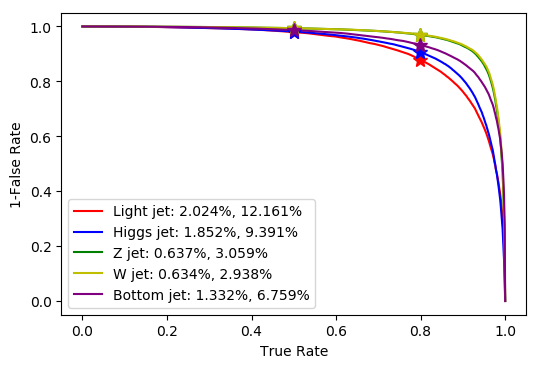}
		\includegraphics[width=0.45 \linewidth]{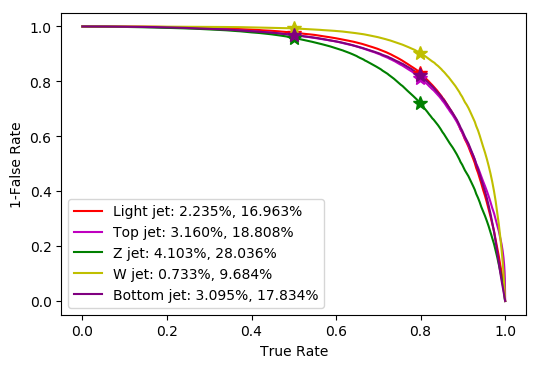}
		\includegraphics[width=0.45 \linewidth]{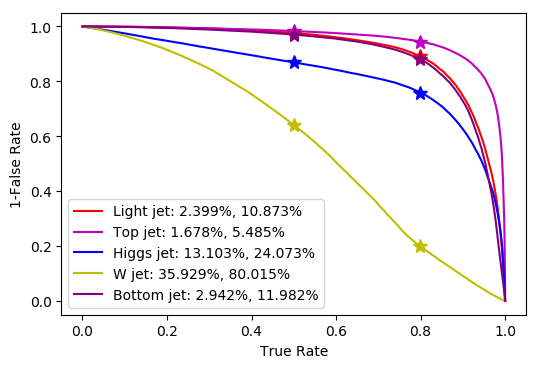}
		\includegraphics[width=0.45 \linewidth]{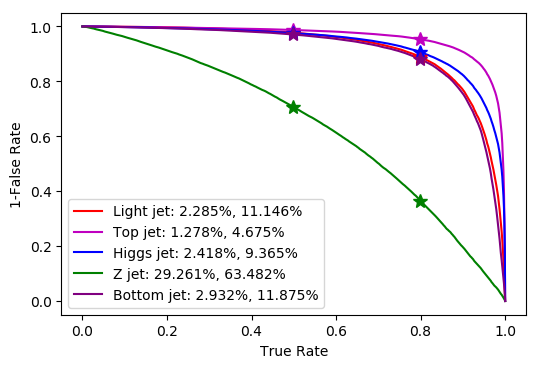}
	\end{center}
	\caption{Jet tagging ROC curve for the second region selection $800 \GeV < p^{fat jet}_{T}$.}
	\label{fig:jet_tagger_800}
\end{figure}

\begin{figure}[t!]
	\begin{center}
		\includegraphics[width=0.45 \linewidth]{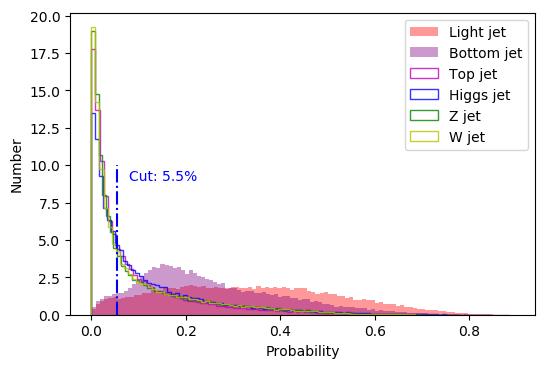}
		\includegraphics[width=0.45 \linewidth]{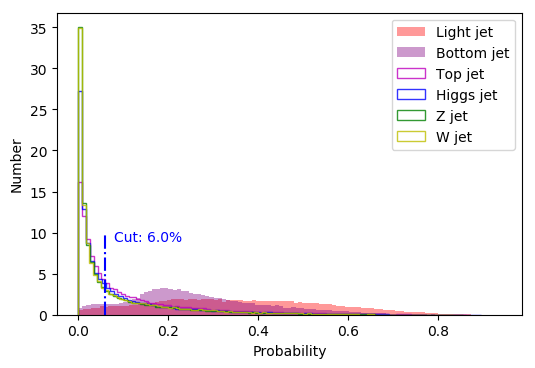}
	\end{center}
	\caption{Probability distribution to identify a jet as light, for $400 \GeV < p^{jet}_{T} < 800\GeV$  (left) and $800 \GeV < p^{jet}_{T}$ (right). The blue dashed line is the cut selected to reject 95\% of the QCD jets (light jets plus bottom jets).}
	\label{fig:non_qcd}
\end{figure}

An important task for the tagger is to reduce QCD contamination in samples of jets from massive particle decays. The performance of the CNN tagger is illustrated in Fig.~\ref{fig:non_qcd}, which shows the distribution of probabilities that a given jet is ``light", for jets in each of the six SM samples. The QCD contamination can be reduced by 95\% with cuts shown by blue lines in the figure, while maintaining high efficiency to tag a massive-object jet in all cases.

\section{Random Forest Classifier}
\label{App:RF}

The key goal of the analyses described in Section~4 is to discriminate between signal events (due to top partner production and decays) and background events (due to SM processes). To maximize the use of kiematic information, we make use of the multivarible machine learning technique, Random Forest (RF) Classifier~\cite{Breiman2001}. RF Classifier algorithms are widely used in high-energy physics and other fields. Their well-known advantages include parallelization, which great reduces CPU time consumption; and avoiding overfitting problems. To construct the RF Classifier used in this paper, we used the {\tt Scikit-Learn}~\cite{scikit-learn} software framework. The RF was constructed to have 10,000 estimators, with the other hyper-parameters left at their default values since they do not have high influence on our analysis.  

In the following, we will discuss the application of the RF algorithm to the $(\gamma\gamma)+2j+X$ channel in the LH model; applications to other channels in Section~4 are similar. For this channel, our dataset consist of 4 event samples, one signal and three background ($hjj$,$tth$,$\gamma\gamma j j$). We select 19 kinematic observables, listed below, as inputs to the RF algorithm. The RF combines the 19 inputs in an optimal way, determined via training on labeled signal and background samples, to return a single number between 0 and 1. RF output close to 1 corresponds to signal while RF output close to 0 corresponds to background. 

The input observables chosen for this analysis are as follows:

\begin{enumerate}

\item $m_{t^{'}}$: Mass of the reconstructed top partner candidate, {\it i.e.} invariant mass of the two highest-$p_T$ fat jets; 

\item $H_T$: Scalar sum of the transverse momenta; 

\item $\met$: Missing transverse energy; 

\item $n_{fj}$: Number of fat jets (clustered with cone size $R=1.0$); 

\item $n_j$: Number of jets (clustered with cone size $R=0.4$); 

\item $PT_0$, $PT_1$: Transverse momenta of the leading and second leading (ordered in $p_T$) fat jets;

\item $m_h$: Mass of the reconstructed Higgs boson {\it i.e.} invariant mass of the two highest-$p_T$ photons;

\item $PT_h$: Transverse momentum of the reconstructed Higgs boson;

\item $PT_{\gamma}^0$, $PT_{\gamma}^1$: Transverse momenta of the leading and second-leading (ordered in $p_T$) photons;

\item $dr_{\gamma\gamma}$: Angular distance between the highest-$p_T$ two photons; 

\item $\frac{E_{\gamma}^0}{E_h}$, $\frac{E_{\gamma}^1}{E_h}$: Ratios of the leading/second-leading photon energies and the Higgs boson energy;

\item $n_b$: Number of tagged $b$-jets;
 
\item $P_t^0$, $P_t^1$: Probabilities that the leading/second-leading fat jet is a top jet; 
 
\item $P_j^0$, $P_j^1$: Probabilities that the leading/second-leading fat jet is a light jet. 

\end{enumerate}

The last four probabilities are provided by the CNN top tagger, see Appendix~\ref{App:JetTag}. Note that not all inputs are linearly independent: for example, $\frac{E_{\gamma}^0}{E_h}+\frac{E_{\gamma}^1}{E_h}=1$. This does not cause problems, since at worst, the RF algorithm will ignore the extra inputs. However, we find that in practice providing such extra inputs explicitly leads to improved performance, presumably because the algorithm can simply choose the best linear combination in each case without having to ``invent" it during training.     

\begin{figure}[t!]
	\begin{center}
		\includegraphics[width=0.6 \linewidth]{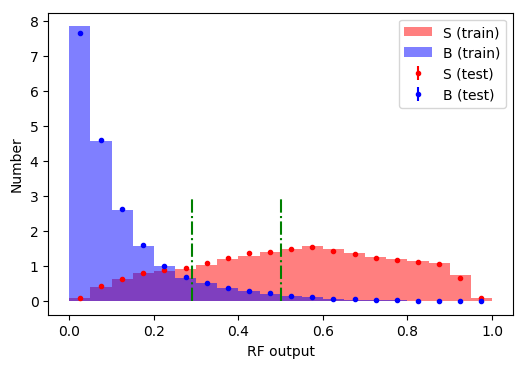}
	\end{center}
	\vskip-0.8cm
	\caption{RF output for the training (bars) and testing (points) samples. The doted-dashed green lines correspond to ``RF I" and ``RF II" signal bins, see Table~\ref{tab:lh_qhtp}.}
	\label{fig:lh_overfitting_test} 
\end{figure}

To demonstrate that the RF classifier is not overfiting, we divide the event samples into ``training" and ``testing" sets, and compare the RF output between the two sets. In~\fig{lh_overfitting_test}, the red/blue bars represent the RF output on training signal/background samples, while the red/blue dots correspond to the testing signal/background samples. The training and testing set RF outputs are in excellent agreement. We conclude that the predictions made by the RF classifier are consistent and the RF model is not overfiting.

\begin{figure}[t!]
	\begin{center}
		\includegraphics[width=0.6 \linewidth]{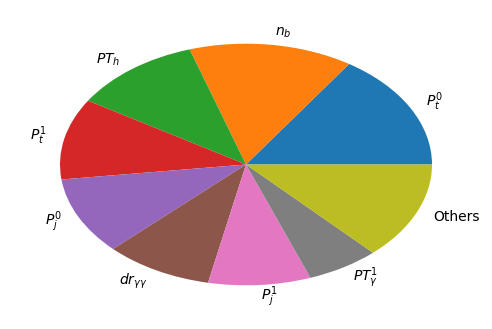}
	\end{center}
\vskip-0.8cm
	\caption{RF inputs classified by their Gini importance. The ``others" slice includes the 11 variables not shown explicitly.}
	\label{fig:feautre_importance} 
\end{figure}

\begin{table}[t!]
	\centering
	\begin{tabular}{|c|c|c|c|c|c|c|}
		\hline
		& Event Number & RF I & RF II & $\sigma$ & Sign $\sigma$  \\ \hline
		signal &$2.8\times10^3$ &$1650$ &$643$ &\multirow{4}{*}{$37$} &\multirow{4}{*}{$4.0$} \\ \cline{1-4}
		$hjj$ &$1.2\times10^4$ &$1017$ &1979 & &  \\ \cline{1-4}
		$tth$ &$6.3\times10^3$ &$1046$ &1064 & &  \\ \cline{1-4}
		$aajj$ &$4.8\times10^4$ &$714$ &2727 & &  \\ \hline		
	\end{tabular}
	\caption{Same as Table~\ref{tab:lh_qhtp}, but with a simpler RF with only 8 input observables.}
	\label{tab:update_lh_qhtp}
\end{table}

To gain some insight into the relative importance of the input observables for signal/background discrimination, we computed the Gini importance for each of the observables. The results are shown in~\fig{feautre_importance}. While no single observable dominates, we note that eight of the observables, shown explicitly in~\fig{feautre_importance}, dominate the analysis. To confirm this, we repeated the significance analysis of Table~\ref{tab:lh_qhtp}, using only the eight leading observables as RF inputs. The results in Table~\ref{tab:update_lh_qhtp} confirm that the significance of signal observation and sign flip discrimination are only modestly degraded compared to the full analysis with 19 inputs.  

\end{appendix}

\bibliographystyle{JHEP}
\bibliography{reference}

\end{document}